\documentclass[]{aa}
\usepackage{txfonts}
\usepackage{graphicx}
\usepackage{natbib}
\bibpunct{(}{)}{;}{a}{}{,}
\def\persec{$\mathrm{s}^{-1}$}
\def\peryr{$\mathrm{yr}^{-1}$}

\begin{document}
\title{Rights and wrongs of the Hipparcos data}
\subtitle{A critical quality assessment of the Hipparcos catalogue}
\author{Floor van Leeuwen% \and Second author\inst{2}% etc
}                     % Do not remove
\offprints{F.van Leeuwen}          % Insert a name or remove this line
\institute{Institute of Astronomy, Cambridge, UK} % \and the second here}
\date{Received: date / Revised version: date}
\abstract{
A critical assessment of the quality of the Hipparcos data, partly supported 
by a completely new analysis of the raw data, is presented with the aim of 
clarifying reliability issues that have surfaced since the publication 
of the Hipparcos catalogue in 1997. A number of defects in the data are 
identified, such as scan-phase discontinuities and effects of external hits.  
These defects can be repaired when re-reducing the raw data. Instabilities
in the great-circle reduction process are recognised and identified in 
a number of data sets. These resulted mainly from the difficult observing 
conditions imposed by the anomalous orbit of the satellite. The stability 
of the basic angle over the mission is confirmed, but the connectivity 
between the two fields 
of view has been less than optimal for some parts of the sky. Both are 
fundamental conditions for producing absolute parallaxes. Although there 
is clear room for improvement of the Hipparcos data, the catalogue as 
published remains generally reliable within the quoted accuracies. Some of
the findings presented here are also relevant for the forthcoming Gaia 
mission.
}
\maketitle
\section{Introduction}
\label{intro}
\subsection{Background}
The Hipparcos mission \citep{esa97,perry97L,fvl97,koval98} involved a wide 
range of new concepts applied under what accidentally became very difficult 
conditions. 
These new concepts were ultimately all aimed at obtaining absolute parallaxes
for a selection of approximately 118000 stars and in this the mission 
largely succeeded. The publication in May 1997 of the catalogue, based 
on the data reduction results obtained by the two independent reduction 
consortia FAST \citep{koval92} and NDAC \citep{lindeg92}, took place according 
to a time schedule set by the Hipparcos Science Team in agreement with the 
European Space Agency \citep{perry97L}. This did not allow for final 
iterations and reflections on methods 
used and results obtained to take place. It is therefore not surprising that 
a number of incidents in the data have not been fully appreciated during the 
preparation of the catalogue. 

Some, but not all, of the problems in the Hipparcos data that have since
been detected can be traced back to the orbit anomaly suffered by the mission. 
Instead of operating at a geostationary position, Hipparcos described a 
geostationary transfer orbit of 10.7~hours, for which the perigee height had 
been increased to an average of 450~km \citep{paper1}. With a rotation 
period of just over 2~hours, a single orbit covered close to five revolutions 
of the satellite. No observations could be made when the satellite was out 
sight for all of the three ground stations. This always happened for about 
2~hours around the perigee passage but could also occur for other reasons.
The data obtained during one orbit will be referred to as a data set.

Since the publication of the Hipparcos catalogue, it has appeared that for 
(small groups of) bright stars one of the two basic conditions for obtaining 
absolute parallaxes may not always have been properly fulfilled. These two 
conditions are:
\begin{itemize}
\item A very stable basic angle, with variations well below 1~mas over a 
10~hour period: all available data show that this condition was well 
fulfilled for more than 99~per~cent of all data sets; 
\item A reconstruction of the satellite's scan phase (the so-called along-scan 
attitude reconstruction) which is at all times based on significant 
contributions from data obtained in both fields of view. The along-scan 
attitude reconstruction provides the reference frame for the Hipparcos
measurements, which come in the form of transit times or abscissae
(the along-scan coordinate of the star \citep[see further Volume 3,][]{esa97}).
\end{itemize}
The latter condition will be referred to as the ``connectivity requirement''. 
It lies at the heart of the Hipparcos concept: through fulfilling this
requirement the effects of parallax displacements in the measurements 
can ultimately (through iterations with the sphere reconstruction and the 
astrometric-parameter determinations) be eliminated from the along-scan 
attitude reconstruction. A reference frame unaffected by parallax displacements
is essential for obtaining absolute parallaxes.

A local failure in the connectivity requirement could explain the apparent 
problem with the Pleiades parallax as derived from the Hipparcos catalogue.
That there may be problems with the Hipparcos determination of the
Pleiades parallax at 8.35~mas \citep{fvl99a,robic99} has been suggested by 
several authors before \citep{pinso98,soder98,reid99,pinso00,stello01,pinso03},
but often without sufficient understanding of how such a discrepancy could 
occur locally only \citep{naray99}. Statistical tests on the data rule out 
global discrepancies in both the parallax zero point and the formal errors 
on the astrometric parameters \citep{areno95,llind95}. 
Independent geometric determinations of the Pleiades parallax are very hard
to obtain, as the cluster velocity  and distance make it unsuitable for 
a convergent-point determination \citep{fvl05}. As not all studies seemed to 
contradict the Hipparcos result \citep{castel02,perciv03}, it remained 
unclear where the fault may lie. In a more recent paper though, Percival
et al.\ considered the traditional longer distance the more probable
\citep{perciv05}.

Some further information has recently come from studies of individual
binary stars in the cluster \citep{pan04,munar04,zwahl04}, which all
indicated the larger distance of 130 to 135~pc. The first two of these 
studies still contain a model element (they are not purely geometric), only
the third of these studies provides a pure geometric determination. All
three provide the distance of a single star in a cluster with a 
diameter of 20~pc \citep{fvl80}, although as bright stars and binaries they
are more likely to be found close to the centre of the cluster
\citep{fvl80,raboud98}. The measurements of the parallaxes of a few 
stars with the HST \citep{soder04} also confirm the larger distance, but 
those observations pertain to differential rather than absolute astrometry.

Limited corrections of the published Hipparcos data are possible and had been
foreseen in its preparation by making available intermediate astrometric and 
photometric data, and providing details of the colour indices used in the 
data reductions. This made possible, for example, the improvements
to the astrometry for the very red, large-amplitude variables 
\citep{pourb00,pourb03,knapp01,knapp03}. The repair of potential 
connectivity problems, however, requires information that can only be 
obtained from the original, raw data.

The present paper does not intend to settle the question of the Pleiades 
parallax, but rather to give a frank overview of issues that have surfaced in 
a new analysis of the raw data as well as in the intermediate data products 
made available with the Hipparcos catalogue \citep{esa97}. As explained above, 
one of these issues, connectivity, may well have a bearing on the determination
of the Pleiades parallax. The present paper is also in a way a justification 
for the preparation and presentation of a complete new reduction of the raw 
Hipparcos data, presented in the accompanying paper. To present a complete new 
reduction of those data involves a very substantial amount of work, which 
can only be justified if it provides very significant improvements in 
accuracy and overall reliability. As will be explained in this and the 
accompanying paper, this is likely to be the case.

\subsection{Structure of the present paper}
This paper is organised as follows. Section~\ref{sec:scan} presents
issues affecting the reconstruction of the along-scan attitude: methods applied
to the reconstruction, and peculiarities of the scanning of the satellite.
Although recognised to exist in the original reductions, the frequency and 
effects of scan-phase discontinuities and external hits had been grossly 
underestimated, and no measures were available to systematically identify
and incorporate these events in the along-scan attitude reconstruction models. 

A major role in the original reductions was played by a process referred to as 
the great-circle reduction \citep{vdmar88,vdmar92,esa97,fvl97}. This process 
allowed, through combining the corrections to abscissae, the along-scan 
attitude and a set of instrument parameters, to obtain a reliable attitude 
reconstruction even in the presence of relatively high errors on the 
reference positions for the stars involved. A single great-circle solution 
used the data obtained over one orbit, and as such involved data for up to 
four revolutions of the satellite. All data obtained over this interval were 
projected on a reference great circle to provide a single abscissa measurement 
for each star observed. These combined abscissae from the great-circle 
solutions will be referred to as orbit-transits. They combine data 
for on average 3 to 4 transits through either of the two fields of view.
Section~\ref{sec:GCR} presents problems experienced in the great-circle 
reduction, most of which were indirectly caused by the orbit anomaly.
When, due to ground-station unavailability or problems with on-board attitude
control, the usable data obtained over an obit covered two revolutions or 
less, the basic angle between the two fields of view (58\degr) allowed for 
instabilities to develop between the estimates of the along-scan attitude 
and the determination of the abscissa corrections. This could also happen 
in orbits where the observations were frequently interrupted by Earth 
occultations (the image of the Earth coming too close to one of the 
apertures to continue observing).
 
Section~\ref{sec:BA} reviews the status of the basic-angle stability. Here
we draw mainly from the results of the new reduction to show that
the stability of the basic angle has been very good, down to a 0.1~mas
level over a 10~hour period. However, 18 orbits (out of 2300) are identified 
where the basic angle did show a significant drift; in almost all cases these
drifts are directly related to known anomalies in the operations of the
spacecraft and payload. 

Section~\ref{sec:conn} presents our current understanding of the importance 
of connectivity: the way data from the two fields of view become connected 
in the data reductions through the reconstruction of the along-scan
attitude. As was stated above, good connectivity and a highly 
stable basic angle are essential for obtaining absolute parallaxes. We also 
examine methods suggested for correcting poor connectivity by 
\citet{makar02,makar03}. 

In Section~\ref{sec:merg} some statistical properties of the abscissa data 
in the published catalogue and the role played by the merging of the data 
from the two reduction consortia are reflected upon. The verification 
possibilities of the formal errors and the parallax zero point for the 
published data are also briefly reviewed in Section~\ref{sec:stat_astr}. 
Finally, Section~\ref{sec:concl} summarises the conclusions of the current 
study.

\subsection{Presentation}
Formal errors on abscissa measurements presented here are generally examined
as a function of the total photon count or integrated intensity of the 
underlying observations. The observing strategy for Hipparcos led to 
variable amounts of integration time being assigned to observations, depending
on stellar brightness as well as on the presence of other program stars
on the main detector area. The integrated intensity is therefore a much more
representative quantity for accuracies than is the stellar magnitude.

Data sets will generally be referred to by their orbit number. Orbits were
numbered sequentially as they took place, irrespective of whether data 
was collected. The first orbit formed part of the test phase and took place
on 5 November 1989, while the actual survey began at orbit 48 on 26 November 
1989. The last orbit with data, orbit 2769, took place 1207.8 days later, on 
18 April 1993.

The abbreviation ``mas'' will be used for 10$^{-3}$ seconds of arc.
Torques will often be represented in the equivalent units for acceleration
(mas~s$^{-2}$) rather than in Nm, with values directly related through the
diagonal elements in the inertia tensor. In integrations for the dynamical 
modelling, however, all calculations are done using the full inertia tensor
and torques expressed in the proper units.

\subsection{Other information}
It is in no way the intention of the current study to discredit the 
work and products of the two data reduction consortia. Much more powerful
computing hardware makes investigations possible that were quite difficult 10 
to 15 years ago. Processing power, data handling facilities and graphical 
display capabilities have all improved very significantly over those years.

This study has progressed over 7 years, starting from the basic ideas of the 
FAST and NDAC consortia on the reduction of the data, and some elementary 
parts of the new reduction are still identical to what was developed 20 
years ago. It shouldn't harm anyone to have a critical look at what was done 
then, to see if better results may be obtained from the same data. The 
Hipparcos data can never be replaced, and if more accurate and reliable 
information can be extracted from it, then this has to be done, and can only 
be done by studying in great detail the raw data and intermediate results 
from the published data. The stars most likely to be significantly affected 
by any potential improvements of the Hipparcos astrometric data are the 
brightest stars (V$<8$), for which the main contributions to the formal errors 
on the astrometric results in the published catalogue are modelling errors 
rather than photon noise. New missions, like Gaia, will find it difficult or 
impossible to obtain data for these stars, for which the images may be 
fully saturated.
\section{Scanning problems}
\label{sec:scan}
Problems related to scanning naturally split in two components: the 
characteristics of the actual scanning motion of the satellite, and our 
modelling of 
these characteristics. The second part of the problem is generally referred to
as the attitude reconstruction. In the Hipparcos reductions little attention 
was given to the first part: it was generally considered too complex for 
modelling, and exact cubic splines were used to fit positional displacements
for the attitude modelling, irrespective of the dynamics of the satellite. 
The only exceptions are the preliminary studies on the torques affecting the 
satellite by \citet{fvl92b} and \citet{fvl97} as part of the NDAC activities.
In the following sub-sections a brief summary of steps taken in the attitude 
reconstruction is presented, followed by descriptions of the three main 
disturbances observed in the scanning motion of the satellite: the scan-phase 
discontinuities, the external hits and the torque irregularities. A more 
detailed discussion of the methods used in the attitude reconstruction can 
be found in Volume 3 of \citet{esa97} and \citet{fvl97}.
\subsection{Steps in the along-scan attitude reconstruction}
\label{sec:att}
\begin{figure}
\centerline{\includegraphics[width=8.5cm]{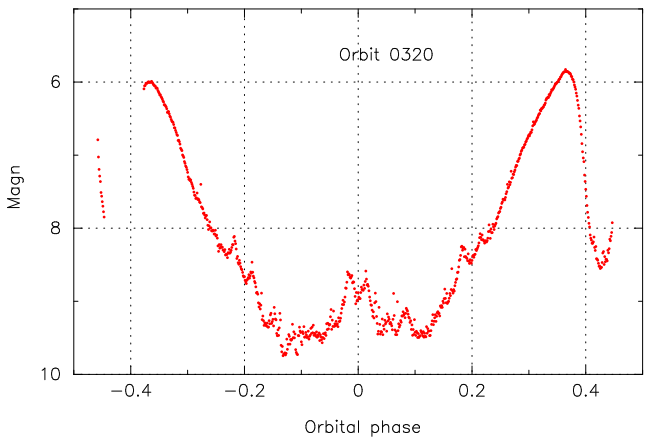}}
\caption[Star mapper background]{The background in the V channel of the 
star mapper for orbit 320 (end March 1990). The vertical scale is given as an 
approximate equivalent of stellar magnitude. The horizontal scale goes from
perigee at -0.5 to apogee at 0.0 to perigee at 0.5. One full orbit covers
close to 5 revolutions of the satellite. Visible near the centre of the 
graph are the crossings of the two fields of view through the zodiacal light 
and the galactic plane. These crossings are repeated around phases -0.2 and 
+0.2.}
\label{fig:sm_back}
\end{figure}
The Hipparcos attitude reconstruction consisted of two main steps: a first
approximation for all three coordinates using the star mapper data
\citep{donat92,fvl92a}, followed by an improvement in the along-scan
attitude based on the main detector measurements as part of the great-circle
reduction \citep{vdmar88,vdmar92}. The first stage of the attitude 
reconstruction had two purposes: to provide the instantaneous position of 
the spin axis of the 
satellite to a $1\sigma$ accuracy of 100~mas, and the along-scan
attitude with a similar or better accuracy. The first was set
by the great-circle solution (see Section~\ref{sec:GCR}) and was based
on the noise introduced when projecting transits on a reference great
circle. The second requirement was essential for the proper identification
of a star position on the modulating grid during a transit. The
modulating grid had a period, as projected on the sky, of 1.2074~arcsec,
and excessively large errors in the first step of the attitude reconstruction
could lead to slit ambiguities.

The first stage of the attitude reconstruction, using the star mapper data, 
could at times be badly affected by high detector noise, in particular
during transits of the Van Allen Belts, which took place twice every orbit
(Fig.~\ref{fig:sm_back}). The background signal during these transits often 
left all but the brightest stars obscured, thereby limiting severely the 
possibility to accurately reconstruct the attitude. Away from those regions, 
the $1\sigma$ errors on the reconstructed position of the spin axis were 
around 80~mas, and for the along-scan phase estimates around $\approx40$ mas, 
but it is likely that for periods of high background the errors were 
considerably larger.

In the second step of the attitude reconstruction the differences between
the predicted and observed positions for the main-grid transits were fitted
using cubic splines,
while at the same time estimating the corrections to the predicted
positions. It is from graphs of these differences that scan-phase
discontinuities and external hits can be recognised. 

\subsection{Scan-phase discontinuities}
\label{sec:jumps}
\begin{figure}
\centerline{\includegraphics[width=8.5cm]{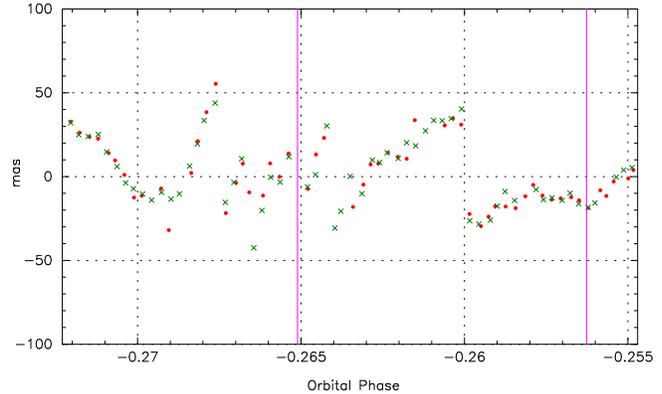}}
\caption[Scan-phase discontinuities]{Scan-phase discontinuities are observed
when the satellite is warming up or cooling down rapidly, for example after 
an eclipse as shown here for orbit 253 (Febr.1990). The discontinuities 
reflect discrete adjustments of the structure of the satellite to the 
changes in temperature. Here 4 jumps (at phases -0.2676, -0.2667, -0.2640,
-0.2600) of around 40 to 70~mas in scan phase are observed from the 
abscissa residuals with respect to the star-mapper based reconstruction
of the along-scan attitude. The crosses and circles represent data from the 
two fields of view. The data points are weighted averages over 10.6~s of 
observations. The vertical lines represent thruster firings.}
\label{fig:jumps}
\end{figure}
Scan-phase discontinuities are briefly explained in Volume 2 (Section 11.4) 
of \citet{esa97}, where their effect on the data was considered insignificant,
as at that stage only very few had been detected.
In the new reduction of the data, however, some 1500 discontinuities ranging 
from 20 to 120~mas have been identified as the main source behind a 
non-Gaussian noise component in the abscissa residuals.

\begin{figure}
\centerline{\includegraphics[width=8.5cm]{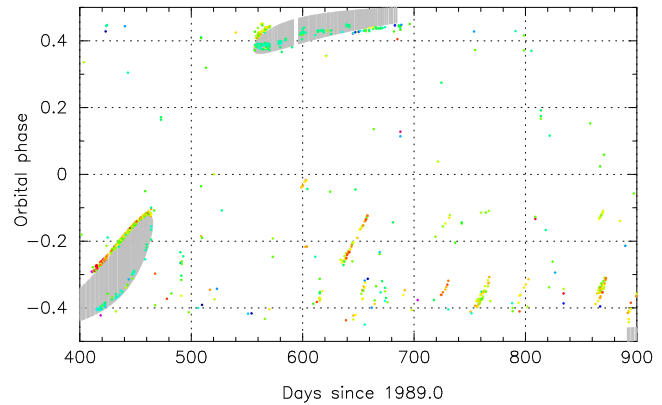}}
\caption[Jumps distribution]{The distribution of scan-phase discontinuities 
over 500 days of the mission. Every dot represents a recorded jump. The grey
areas are times of eclipses. Jumps are clearly concentrated towards the start
and end of eclipses, but also occur systematically away from those events. As 
the orbital period of the satellite was close to five complete rotations, the 
rotation phase of the satellite for a given orbital phase changed little
from one orbit to the next. Jumps tend to concentrate around a fixed 
rotation phase of the satellite, causing the high levels of correlation in the
diagram.}
\label{fig:jumps_dist}
\end{figure}

A scan-phase discontinuity is like a jump (Fig.~\ref{fig:jumps}), and as such 
is directly related to the rigidity of the satellite: as the total angular 
momentum of the satellite is not affected, a jump has to be the result of 
one part of the satellite moving with respect to some other part. A clear 
indication of which part may be moving comes from the temporal distribution 
at which jumps take place: a large fraction is directly linked to eclipses, 
when the satellite moves in to, or out of, the Earth shadow. These jumps 
appear to be immediate reactions to the strong temperature changes taking 
place at these instances (Fig.~\ref{fig:jumps_dist}). Jumps are otherwise 
found to be concentrated around (but not restricted to) an interval of about 
$45\degr\pm25\degr$ in the rotation phase of the satellite. The rotation phase,
by definition, represents the direction of the solar radiation seen from the 
satellite. Both aspects point to an external element of the satellite, and the 
most obvious candidate is any one of the three solar panels, as was also 
identified in the earlier preliminary investigations on this subject. 

We can estimate the amplitudes of the possible displacements from the sizes,
masses and inertia moments involved. Each solar panel measures 1.69 by 1.39~m
and has a mass of 6.363~kg. They are connected (on the shorter side) to the 
spacecraft at 0.9546~m from the spin axis. Assuming an even mass distribution, 
the inertia moment of a single solar panel around the spin axis is 
$23.15~\mathrm{kg~m^2}$. The inertia moment around the spin axis for the 
entire satellite is $459~\mathrm{kg~m^2}$, giving
a ratio of $1:19$ for a single solar panel versus the rest of the spacecraft.
The rotational displacement of a solar panel required to give a 40~mas
rotation of the payload is therefore just over 0.75~arcsec. At the point
of attachment of the solar panels, a 0.75~arcsec rotation amounts to a shift 
of 3.4~$\mu$m. Thus, very small discrete displacements in the solar panel 
positions with respect to the spacecraft are sufficient to cause the 
observed scan-phase discontinuities.

Jumps tend to be negative at the start of an eclipse, and positive after,
which may be expected when heating up and cooling down cause opposite 
movements. Jumps do occur also in a systematic manner away from eclipses,
and are noticeably more frequent in the few hours after the perigee passage
(Fig.~\ref{fig:jumps_dist}).
During the perigee passage the outer layers of the Earth atmosphere can affect
the temperature of the solar panels, and possibly in some cases even the
position.

\begin{figure}
\centerline{\includegraphics[width=7.8cm]{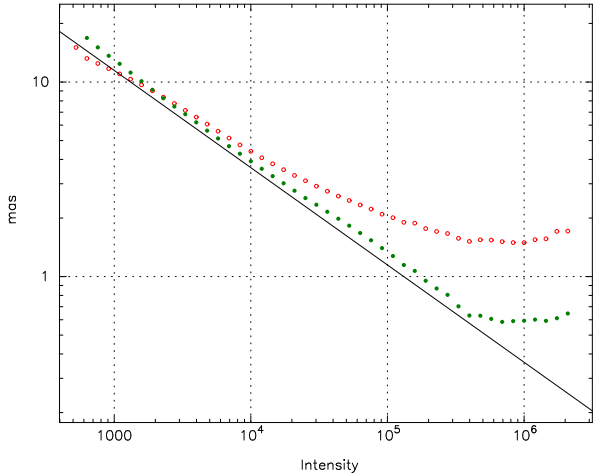}}
\caption[Old abscissa dispersions]{The reduction in the abscissa dispersions
due to, amongst others, the detection of scan-phase jumps shows in a comparison
between results before removal (open symbols) and at the end of the fourth
iteration (filled symbols), when nearly all these defects had been detected 
and incorporated in
the modelling. A further difference is visible for the low intensities, where 
in the earlier stages of the reduction the cut-off criterion for poor-quality
measurements had been set too low, creating artificially low dispersions.
The slope of the diagonal line represents the contribution of the 
Poisson-noise.}
\label{fig:absc_old}
\end{figure}
The overall effect of phase jumps on the published data is difficult to 
assess. The amount of data potentially directly affected can be estimated 
from the number of jumps detected: around 1500. The time interval around a jump
for which the attitude reconstruction is affected if the presence of the
jump is not incorporated in the attitude model is about 150~s. The total
amounts to approximately 3 to 4 per~cent of the mission. It is difficult
to estimate how these jumps may have affected the along-scan attitude
reconstruction in the great-circle reduction due to the way the data are
projected on a reference great circle. As a result, local problems in the 
scanning can replicate between different revolutions of the satellite and
between different parts of the scan. The affected time may therefore be
much larger, with the effects of a phase jump repeating at intervals of
the basic angle, 58\degr. Such error correlations are observed in the 
statistics for abscissa residuals \citep{vLDWE} for intervals up to 
eight times the basic angle. This ``spreading out'' of scan-phase problems
is for example displayed by the systematic effects observed in the abscissa 
residuals in orbits 721 to 796 (covering days 638 to 660 in 
Fig.~\ref{fig:jumps_dist}). These orbits are all affected 
by phase jumps that are not directly associated with an eclipse. The 
abscissa residuals for these orbits in both the FAST and NDAC solutions show
a relatively high level of systematics, spread over all rotation phases, 
and in some cases resembling phase jumps. 
Contrary to the assumptions for the published data, statistics from the new 
reduction clearly indicate that these phase jumps did impose a significant, 
non-Gaussian, noise component on the abscissa residuals. This is shown most 
clearly in the noise levels of the abscissa residuals before and after
detecting and taking care of these events (Fig.~\ref{fig:absc_old}).

\subsection{External hits}
\label{sec:hits}
The first hits of the spacecraft were recognised by the NDAC team during 
initial inspections of the gyro data. A hit of the spacecraft causes a 
discontinuity in its inertial rates. Such discontinuities were also caused
by thruster firings, but the instances of the firings were known from the 
satellite telemetry, and firings always took place in a designated time
interval. Inertial rate discontinuities not associated with thruster firings
and not in the designated time interval could safely be attributed to
external hits. The largest of these caused rate changes at a level of a few
arcsec~\persec\ and four of these have been recorded over the mission. In the
NDAC solution these events were incorporated in the attitude model in the
same way as thruster firings; in the FAST reductions the gyro data were not
analysed and the affected data have most likely been rejected at some 
stage in the reductions.

\begin{figure}
\centerline{\includegraphics[width=8.5cm]{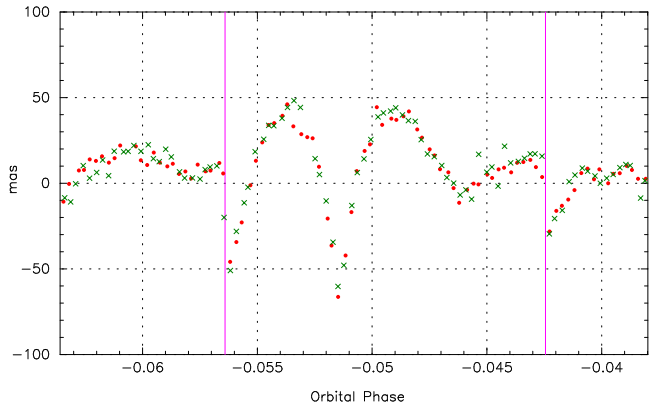}}
\caption[An external hit]{A small external hit of the satellite as reflected
in the abscissa residuals relative to the star-mapper based scan-phase
reconstruction. The data are for orbit 715 (Oct.1990), for which the scan 
velocity at orbital phase -0.05148 changed abruptly by 4.7 mas~\persec.
The crosses and circles represent data from the two fields of view. The 
data points are weighted averages over 10.6~s of observations. The vertical 
lines represent thruster firings.}
\label{fig:hit}
\end{figure}
Many more smaller hits took place, but these could not be recognised from 
the gyro data because of noise levels. Smaller hits do show up, however,
in the behaviour of the abscissa residuals with respect to the star-mapper
based attitude reconstruction, as shown by the example in Fig.~\ref{fig:hit}.
Some 150 of these hits have now been recorded, none of which had been 
incorporated in the reduction of the Hipparcos data. As with the
phase jumps, the fraction of time directly affected is relatively small, 
about 0.4 per~cent, but due to propagation in the great-circle reduction,
more data may have been indirectly affected. 

The actual energy transfer associated with a hit producing a 
$\Delta\omega=5$~mas~\persec\ rate jump is surprisingly small. With an 
inertia moment around the spin axis of 
$\mathrm{\textbf{I}}_{zz}=459~\mathrm{kg~m^2}$ and a nominal scan velocity 
of 168.75~arcsec~\persec, the change in rotational energy is about 9nJ. The 
change in angular momentum provides a lower limit of the typical particle 
mass involved:
\begin{equation}
m \leq \frac{\Delta\omega\mathrm{\textbf{I}}_{zz}}{v\cdot r_\mathrm{max}},
\end{equation} 
where $v$ the velocity of the particle, and $r_\mathrm{max}$ the maximum 
arm-length of the impact position (about 0.95~m). At typical velocities of 
around 30~km~\persec, the masses of the particles involved are small 
($10^{-3}$ to $10^{-4}$~mg). However, the shape of the satellite 
is such that the torque for the spin direction caused by a particle hit 
will have been relatively inefficient in most cases, and the actual particle 
masses involved in these hits are likely to be larger than indicated by the 
angular-momentum transfer for the spin axis. 
The same uncertainty of where a particle may have hit, and how much it 
changed the rates on the other two axes (where due to the 100 times higher
noise levels the measurements are too unreliable), means that we cannot
obtain reliable statistical information from these events. 
\subsection{Torque disturbances}
\label{sec:torq}
\begin{figure}
\centerline{\includegraphics[width=8.5cm]{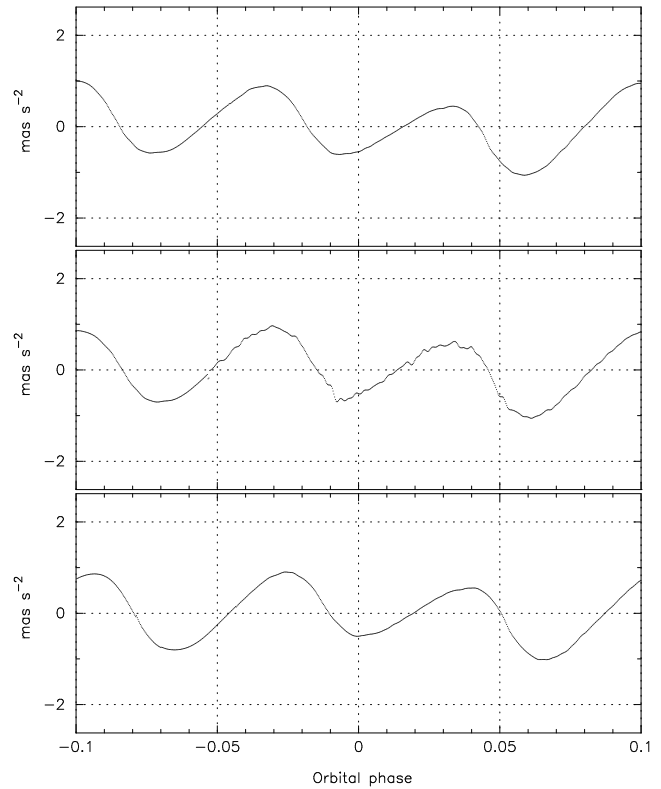}}
\caption[The bumpy road]{Reconstructed torques around the spin axis for one
rotation of the satellite around apogee, expressed in equivalent accelerations.
From top to bottom: orbits 1317, 1318 and 1320. The torques in orbit 1318
are significantly disturbed, possibly due to solar wind following high solar
activity.}
\label{fig:bumpy}
\end{figure}
Torque disturbances are observed around apogee during or shortly after times 
of high solar activity (Fig.~\ref{fig:bumpy}). These may be related to the 
position of the bow-shock of the Earth's magnetic field, which can go down 
to the altitude of geostationary orbits in the presence of strong solar 
winds. The main effect on the data is a requirement for a significant increase 
in the number of nodes in the spline function, which can weaken the solution. 
In the NDAC and FAST data analysis increases in the number of nodes were 
determined automatically on the basis of the dispersion of the data. In
the published data, the data sets concerned do not appear to be badly affected.

Problematic attitude behaviour was also encountered around the start and end
times of eclipses, when solar radiation torques change rapidly. It is
difficult to retrace in the Hipparcos data how the NDAC and FAST solutions
coped with these situations, as the same orbits are also affected by 
the scan-phase discontinuities. In both reduction chains the abscissa residuals
for these orbits show well above average levels of abscissa-error correlations.

\subsection{Conclusions on scanning problems}
\begin{figure}
\centerline{\includegraphics[width=8.5cm]{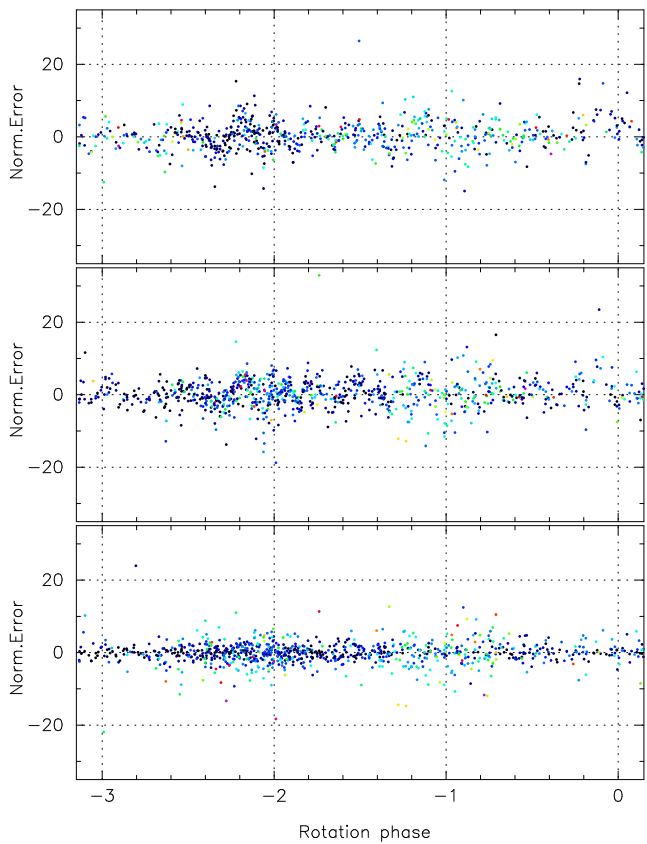}}
\caption[Correlated errors]{Abscissa residuals for part of orbit 237. From 
top to bottom: results from the FAST, NDAC, and new reductions. The residuals 
in the FAST and NDAC solutions are clearly correlated, which may be a 
reflection of the presence of uncorrected scan phase discontinuities in 
this orbit.}
\label{fig:corr}
\end{figure}
A re-analysis shows that scanning problems affect the quality of the Hipparcos 
data for a significant number of orbits. The great-circle 
reduction procedure tended to transpose any such problems to other parts of 
the great-circle solution due to the simultaneous solution for the abscissa 
corrections and the along-scan attitude modelling (see Section~\ref{sec:GCR}). 

One aspect of the systematics in the abscissa residuals is the way 
these are often correlated between the FAST and NDAC solutions, as shown 
for example in Fig.~\ref{fig:corr}. These correlations had also been noticed 
by Fr{\'e}\-d{\'e}ric Arenou \citep[see Volume 3, Chapter 17, of ][]{esa97}.
They could be expected in two situations: when there are (common) systematic 
errors in the astrometric catalogues used for the final reductions, or when 
the errors are caused by intrinsic, uncorrected problems with the data, such 
as scan-phase or velocity jumps. The first solution seems unlikely (but can't 
be entirely excluded) due to the independent developments of intermediate 
catalogues used in the reductions. The second solution is more plausible, 
and could be the result of the scanning problems (scan-phase discontinuities, 
hits) described above. In both cases there is clear room for improvements in 
the Hipparcos results. 

\section{The great-circle reduction}
\label{sec:GCR}
\subsection{Brief overview}
\label{sec:overv}
\begin{figure}
\centerline{\includegraphics[width=8cm]{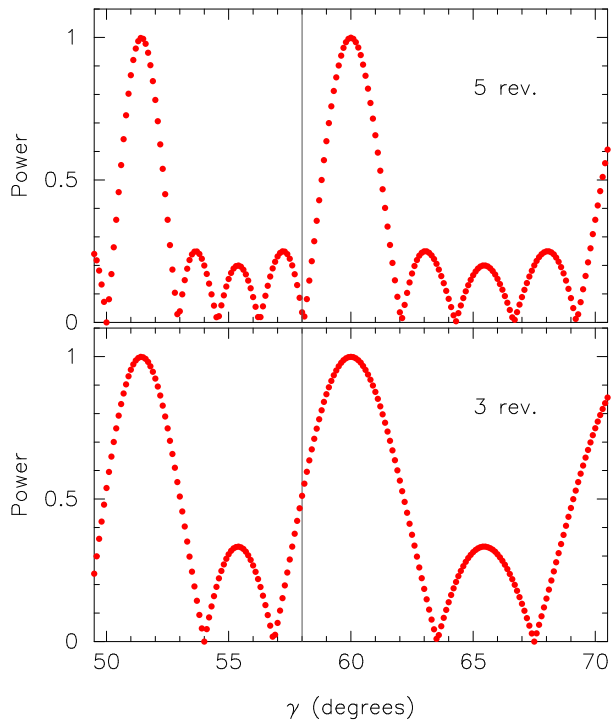}}
\caption[Basic angle selection]{Resonance tests (see text) on basic 
angle values
for 5 (top) and 3 (bottom) revolutions of the satellite. The value of 
58\degr\ at 5 revolutions was chosen, but this allowed some resonances 
in the recovery mission, where the lengths of data sets are rarely more 
than 4 revolutions.}
\label{fig:BAsel}
\end{figure}
For an overview of the mechanism of the great-circle reduction the reader
is referred to \citet{fvl97} or Volume 3 of \citet{esa97}. Here only the
main points will be reviewed.

The great-circle reduction (GCR) combined the solutions of three parameter
sets: corrections to the along-scan attitude, the stellar abscissae, and
the instrument parameters. These solutions were linked to accommodate the
rather large errors in the \textit{a priori} astrometric data for the 
mission's program stars. By measuring the same stars several times 
over one orbit the contributions from the three parameter sets could be
disentangled. For the nominal mission the length of time covered
by one great-circle solution would have been 5 revolutions (of 2.13~h) of 
the satellite, resulting in on average 6 to 7 field transits per star.
The basic angle had been optimised at 58\degr\ for this situation. 
Optimisation is obtained through convolving observations of the
two fields of view in successive revolutions of the satellite with a
sinusoidal function with a period equal to the basic angle. This is done
for different numbers of satellite 
revolutions and different basic angles, as shown in Fig.~\ref{fig:BAsel} for 
3 and 5 revolutions: when an integer number of basic angles fits
within an integer number of revolutions of the satellite, a resonance is
created which can lead to instabilities in the astrometric solution.
 
The limitations imposed by the recovery mission provided between 2 and 4 
revolutions, and an average over the mission of 3.5 field transits per star 
per orbit. To see how this may have affected the overall quality of the 
Hipparcos data, the abscissa residuals as provided for each star in the 
Intermediate Astrometric data file of the Hipparcos catalogue were regrouped 
again per orbit and reduction consortium. 
   
\subsection{Short data sets}
\label{sec:short}
\begin{figure}
\centerline{\includegraphics[width=8.5cm]{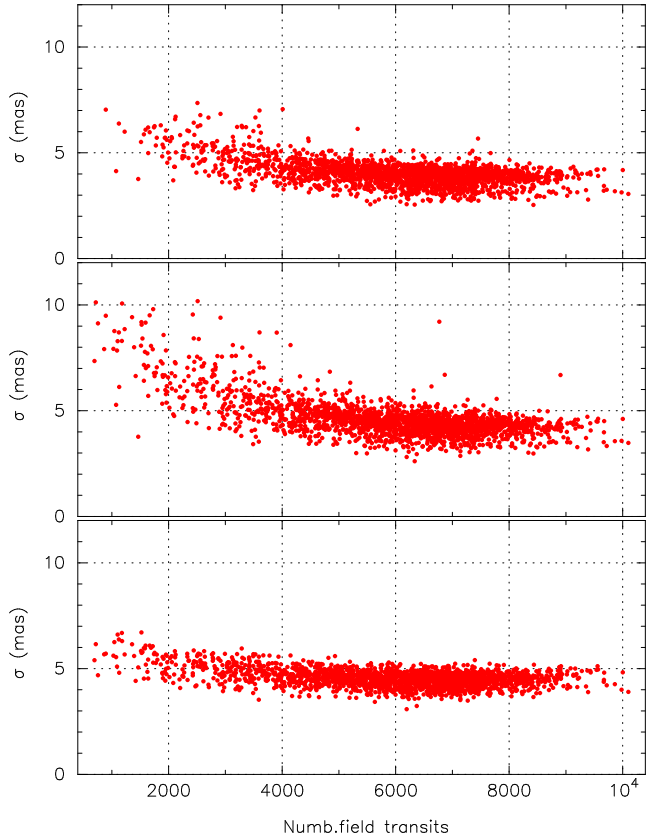}}
\caption[Abscissa dispersions]{The abscissa dispersions for stars brighter
than 8$^\mathrm{th}$ magnitude against the total number of field transits (as 
a measure of the time coverage of the data). Each point represents the data 
for one orbit. From top to bottom: results for FAST, NDAC and the new 
reduction. Increased dispersions for the short data sets (low numbers of field 
transits) are an indication of the instabilities that could occur in the 
great-circle solution. The application of a $3\sigma$ cut-off for outliers 
make the FAST and NDAC data look better than the new reduction, where no 
cut-off is applied other than as indicated by signal-quality criteria.}
\label{fig:AbsDisp}
\end{figure}
\begin{figure}
\centerline{\includegraphics[width=8.5cm]{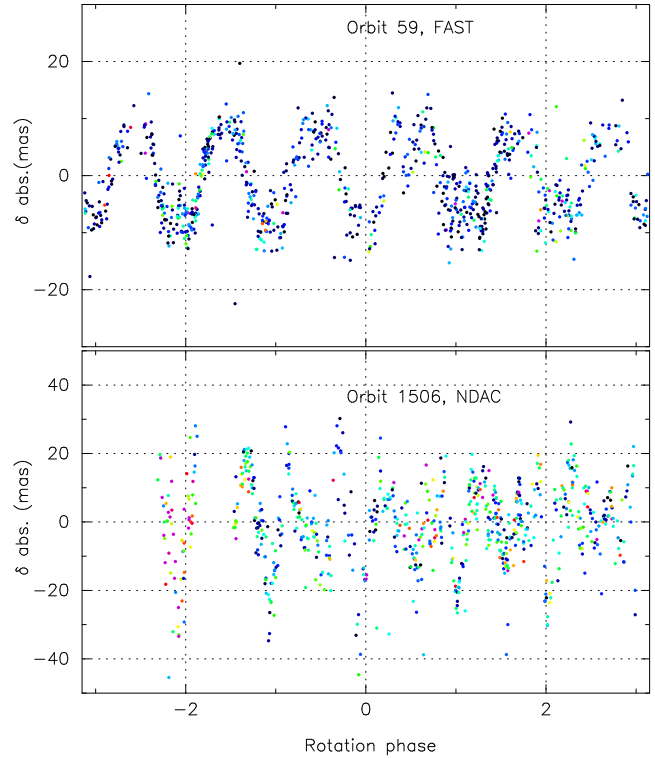}}
\caption[Abscissa residuals or0095]{The abscissa residuals for orbit 95 in 
the three reductions, showing the systematics left by instabilities in the
great-circle reduction as applied by FAST and NDAC.}
\label{fig:AbsRes_95}
\end{figure}
\begin{figure}
\centerline{\includegraphics[width=8.5cm]{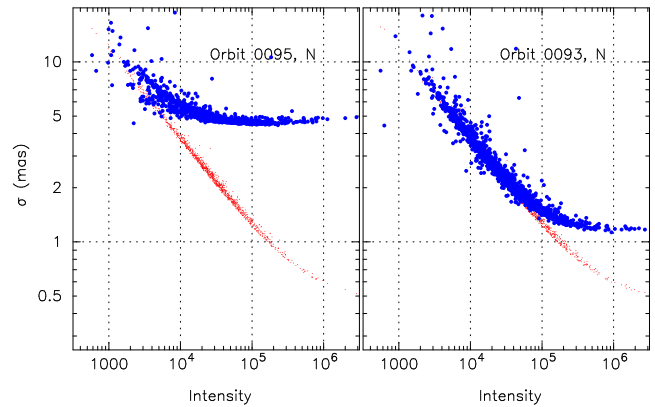}}
\caption[Formal Errors orbits 93 & 95]{Formal errors of abscissa residual 
as a function of total
intensity of the transit for orbits 95 (see also Fig.~\ref{fig:AbsRes_95}) 
and 93 in the NDAC reduction. The thin dots in the background are the 
equivalent values for the new reduction, for which the formal errors are 
given as a function of the total transit photon count or integrated intensity
and the relative modulation amplitude (see Section~\ref{sec:HIC}). 
The systematic errors in orbit 95 were hidden behind large formal 
errors assigned to the abscissa residuals.}
\label{fig:formerr9395}
\end{figure}

An indicator of the global quality of the great-circle reduction results
is the abscissa dispersion for the brighter stars. Figure~\ref{fig:AbsDisp}
shows the results for the FAST, NDAC and new solutions, with a clear increase
in dispersion for the shortest data sets, but also with the majority of the 
data sets performing quite well. In the weighting of the abscissa residuals 
for the astrometric parameter solutions of individual stars, the formal
errors took into account the overall variance of the abscissae in an orbit.
Whenever the variance was above average, however, the abscissa errors
were mostly systematic rather than random. An example of this is shown 
in the results for orbit 95, December 1989, containing two large gaps due
to Earth occultations which prevented the ring in the great-circle reduction
from being closed (Fig.~\ref{fig:AbsRes_95}). The closing of the ring, 
achieved 
by obtaining data well distributed over a great circle, was an essential 
condition to stabilise the solution. The way these systematic errors
have been dealt with in the further application of those data is clear when
examining the formal errors assigned to the abscissa residuals, and comparing 
these with a well-covered orbit (Fig.~\ref{fig:formerr9395}, where the total
transit intensities have been reconstructed from the new reduction). The effect
of this on the final astrometric data is probably small. For the majority of
brighter stars the formal errors assigned to the abscissa residuals in data 
sets with large systematic errors are relatively large so that they added 
little weight to the solutions. The new solution allows a full recovery of all 
data thus affected.

\subsection{The 6$^{th}$ and 12$^{th}$ harmonics}
\label{sec:six}
\begin{figure}
\centerline{\includegraphics[width=8.5cm]{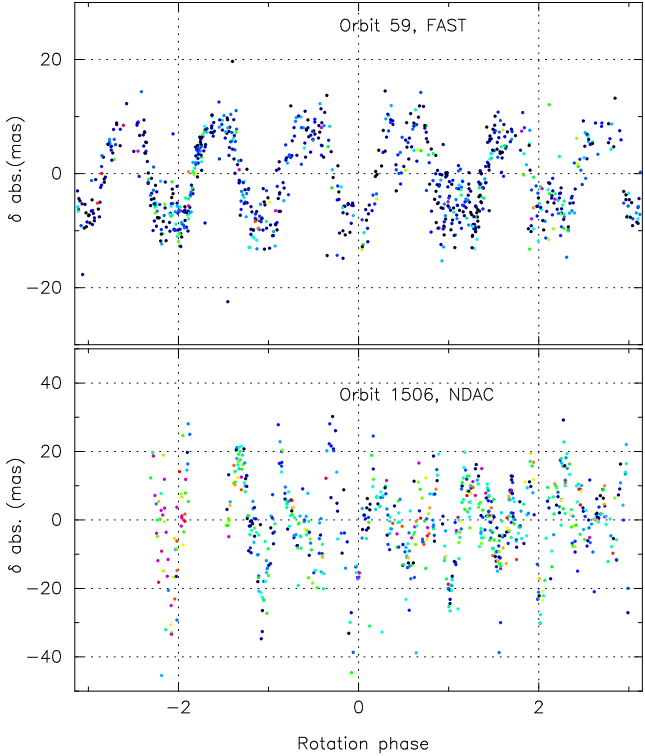}}
\caption[Abscissa residuals for orbit 59]{Top: The abscissa residuals for orbit
59 in the FAST reduction, the only orbit left with a clear 6$^\mathrm{th}$ 
harmonic modulation of the abscissa residuals. Bottom: the same for orbit 
1506 in the NDAC reductions, an extreme example of a 12$^\mathrm{th}$ 
harmonic modulation.}
\label{fig:abs_0059}
\end{figure}
The so-called $6^\mathrm{th}$ harmonic concerned a special kind of 
instability in the great-circle reduction caused by the proximity of 6 
times the basic angle to a full circle. As shown in Fig.~\ref{fig:BAsel}, this 
significantly reduced the stability of shorter data sets. The 
$6^\mathrm{th}$ harmonic has been removed from the data as part of the 
sphere solution, though one data set in the FAST reductions seems to have 
slipped through (orbit 59) and constitutes an example of how data sets could 
be affected (Fig.~\ref{fig:abs_0059}). No other FAST or NDAC data sets appear 
to be affected. However, 12$^\mathrm{th}$ harmonics were left in the data, 
and an extreme example of this is also shown in Fig.~\ref{fig:abs_0059}. 
The occurrence of strong 12$^\mathrm{th}$ harmonics in the abscissa residuals 
is generally restricted to data sets with less than one full revolution. Most 
of these data sets have only been accepted by NDAC, but data sets with quite 
significant 12$^\mathrm{th}$ harmonics do also occur in the FAST reductions.

\subsection{Conclusions on the great-circle reduction induced effects}
\label{sec:concl_GCR}
The coupling of the three parameter solutions in the great-circle reductions
was necessary at the start of the processing because of the relatively
high errors in the initial astrometric parameters for the program stars. 
The special conditions of the mission added vulnerability to instabilities 
to this process, but the vast majority of data sets are largely unaffected. 
Using the astrometric parameters as published in a new reduction shows that 
the solutions of the three parameter sets can now be de-coupled, eliminating 
these instabilities.

\section{The basic angle}
\label{sec:BA}
The stability of the basic angle, the angle between the two fields of view,
constitutes a critical requirement of the Hipparcos mission concept. It is
one of the two fundamental conditions for obtaining a rigid reconstruction 
of the sky and determines a limit on the final accuracies of the astrometric 
parameters. One specific systematic modulation of the basic angle changes 
the zero point of the parallaxes and needs to be avoided above all others.
\subsection{Relation with the parallax zero point}
\label{sec:zero}
\begin{figure}
\includegraphics[width=8cm]{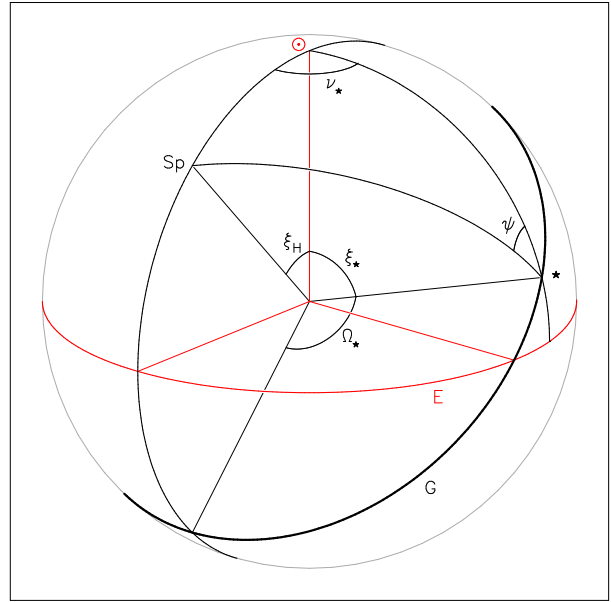}
\caption[Heliotropic reference frame]{A great circle (G) in the 
heliotropic reference frame, showing the relation between the angles 
$\psi$, $\Omega_*$ and $\xi_\mathrm{H}$ for the calculation of 
the parallax displacements along the circle for a star (*). The direction of
the spin axis is indicated by ``Sp''.
}
\label{fig:helio_ref}
\end{figure}
To investigate the relation between a parallax zero-point and a modulation 
of the basic angle we first look at parallax displacements in a heliotropic 
reference frame: a reference frame in which the direction of the Sun is 
fixed (Fig.~\ref{fig:helio_ref}). We assign the North-pole of this reference
system to the direction of the Sun. The displacement resulting from a 
parallax $\varpi$ is now restricted to the meridian, and a function of
the co-latitude $\xi$ only:
\begin{equation}
\Delta\xi(t) = \varpi\sin\xi(t),
\label{equ_hel_par}
\end{equation}
which reaches, as expected, a maximum for $\xi(t)=90$\degr. The time 
dependence reflects the inertial rotation of the heliotropic reference system.
Almost every great
circle in the Hipparcos survey is tilted by 43\degr\ with respect to this
system, the only exceptions are the Sun-pointing data sets, which are tilted 
by 0\degr. In the coordinates of Fig.~\ref{fig:helio_ref} the local
inclination of the great circle with respect to the local meridian is given
by the angle $\psi$:
\begin{equation}
\sin\psi = \frac{\sin\Omega_*\sin\xi_\mathrm{H}}{\sin\xi_*},
\end{equation} 
where $\xi_*$ is the contemporary value of $\xi(t)$ for the star observed,
and $\Omega_*$ the orientation phase or abscissa of the star. Thus,
for Sun-pointing data sets ($\xi_\mathrm{H}=0\degr$) all parallax 
displacements along the circle are zero. In all other cases, the parallax 
displacements along the great circle are given by:
\begin{equation}
{\rm d}a = -\Delta\xi\sin\psi,
\end{equation}
which, together with Eq.~\ref{equ_hel_par} gives:
\begin{equation}
{\rm d}a = -\varpi \sin\Omega_*\sin\xi_H.
\label{eq:pardispl}
\end{equation}
Thus, the parallax displacement along the great circle only depends on the 
abscissa of the star and the instantaneous solar aspect angle of the 
satellite's spin axis ($\xi_\mathrm{H}$), but not on the ordinate $\xi_*$ of 
the star. What is important here is the coefficient $\sin\Omega_*$ in the 
parallax-offset relation. If a similar coefficient can be caused by modulation
of the basic angle, then the zero-point of the parallax system is no longer 
secure.  

To understand the mechanism of the derivation below it is important to
realise how the two processes which, through iterations, ultimately 
determine the Hipparcos catalogue, operate. The first, the along-scan attitude
reconstruction, uses abscissa residuals as observed for stellar transits in 
the two fields of view at the same instance of time (or rotation phase 
$\Omega_H$ of the satellite). The only distinction between the data from the
two fields of view is through a fixed correction (per orbit) to the assumed 
value of the basic angle. The second process, the astrometric parameter 
determination, uses data for the same star as observed in the two fields of 
view at different times. All observations of a star during one orbit are 
associated with a nearly fixed value of the orientation angle $\Omega_*$. 
Every $\Omega_*$ value differs from two associated $\Omega_H$ values by 
$\pm\gamma/2$, and vice versa ($\gamma$ is the basic angle between the two 
fields of view, $\gamma/2\approx 29$ degrees). This determination makes no 
distinction between data from the two fields of view.

We will examine the effects of a modulation of the basic angle by  
$\delta h_0 = A\cdot\cos\Omega_\mathrm{H}$, where $h_0$ is defined as the 
correction to half the basic angle and the subscript ``H'' refers to the 
position of the satellite $x$-axis. This modulation reflects
in the abscissa residuals differently for observations of the
same star in the preceding or following field of view:
\begin{eqnarray}
\mathrm{d}a_\mathrm{p,*}&= \phantom{-}A\cdot\cos\Omega_{H,p} &= \phantom{-}A\cdot\cos(\Omega_*-\gamma/2),\nonumber\\
\mathrm{d}a_\mathrm{f,*}&= -A\cdot\cos\Omega_{H,f} &= -A\cdot\cos(\Omega_*+\gamma/2).
\end{eqnarray}
For the astrometric-parameter determination, the average combined effect from 
the two fields of view is:
\begin{equation}
\mathrm{d}a_* = A\cdot\sin\Omega_*\sin(\gamma/2),
\label{eq_da_ba}
\end{equation}
which is directly correlated with the parallax displacements as given by
Eq.~\ref{eq:pardispl}. Such a modulation of the basic angle will produce a 
systematic offset in the measured parallaxes: on average, the parallax
measurements for all stars will be offset by a fixed amount of:
\begin{equation} 
\varpi_0 \approx -A\frac{\sin(\gamma/2)}{\sin\xi_\mathrm{H}}.  
\end{equation}

A parallax zero-point offset will affect the along-scan attitude 
reconstruction, which in turn affects the measured abscissa residuals. In 
order to estimate a possible equilibrium that may be reached for these 
corrections, an additional ``efficiency'' factor $F$ is included in 
Eq.~\ref{eq_da_ba}:
\begin{equation}
\mathrm{d}a_* = A\cdot\sin\Omega_*\cdot\sin(\gamma/2)\cdot F,
\label{eq_da_ba_a}
\end{equation}
The parallax zero-point offset is similarly affected by this factor:
\begin{equation} 
\varpi_0 \approx -A\frac{\sin(\gamma/2)}{\sin\xi_\mathrm{H}}\cdot F.
\label{eq_pi_ba} 
\end{equation}
The abscissa offsets as a result of the parallax zero point for a 
given value of $\Omega_H$ are (Eq.~\ref{eq:pardispl}):
\begin{eqnarray}
\mathrm{d}a_{p,H} &=& - \varpi_0\sin(\Omega_H+\gamma/2)\sin\xi_H,\nonumber \\
\mathrm{d}a_{f,H} &=& - \varpi_0\sin(\Omega_H-\gamma/2)\sin\xi_H.
\end{eqnarray}
The attitude model uses the average of the abscissa residuals for the 
two fields of view:
\begin{equation}
\mathrm{d}a_H = -\varpi_0\sin\Omega_H\cos(\gamma/2)\sin\xi_H,
\end{equation}
which together with Eq.~\ref{eq_pi_ba} gives:
\begin{equation}
\mathrm{d}a_H = A\sin\Omega_H\sin(\gamma/2)\cos(\gamma/2)\cdot F,
\end{equation} 
which is the attitude modulation caused by the parallax zero point, and
relative to which the abscissa residuals are measured. For a given star,
there will be contributions of this kind from observations in the preceding 
and following fields of view:
\begin{eqnarray}
\mathrm{d}a_{p,*} &=& A\sin(\Omega_*-\gamma/2)\sin(\gamma/2)\cos(\gamma/2)
\cdot F, \nonumber\\
\mathrm{d}a_{f,*} &=& A\sin(\Omega_*+\gamma/2)\sin(\gamma/2)\cos(\gamma/2)
\cdot F.
\end{eqnarray} 
The average of these contributions equals:
\begin{equation}
\mathrm{d}a_* = A\sin\Omega_*\sin(\gamma/2)(\cos(\gamma/2))^2\cdot F.
\label{eq_ba_att}
\end{equation}
Equilibrium can be reached in the iterations when the actual abscissa offsets 
caused by the basic-angle modulation as given by Eq.~\ref{eq_da_ba} equal the 
sum of the contributions from the induced attitude modulation 
(Eq.~\ref{eq_ba_att}) and the measured abscissa residuals 
(Eq.~\ref{eq_da_ba_a}):
\begin{equation}
1 = (1 + (\cos(\gamma/2))^2)\cdot F,
\end{equation} 
where we divided left and right by the common factor
$A\sin\Omega_*\sin(\gamma/2)$.
Thus, $F=(1 + (\cos\gamma/2)^2)^{-1}$, and the parallax zero point becomes:
\begin{equation}
\varpi_0 = -A\frac{\sin(\gamma/2)}{(1+(\cos(\gamma/2))^2)\sin\xi_H}
\approx -0.40A.
\end{equation}
Extensive tests of the zero point of the Hipparcos parallaxes have 
revealed no systematic offset down to an accuracy of 0.1~mas 
\citep{areno95,llind95}, from which we can infer $|A|<0.25$~mas. 
The equivalent factor for the Gaia design is 0.70 (assuming $\gamma=99.4\degr$ 
and $\xi_H=50\degr$), making Gaia more sensitive to such disturbances than
Hipparcos was. 

Any other systematic spin-synchronous modulation will be uncorrelated
with the parallax displacements and will be absorbed in the attitude 
modelling and abscissae noise.
\subsection{Basic-angle evolution, stability and drifts}
\label{sec:ba_evol}
\begin{figure}
\includegraphics[width=8.5cm]{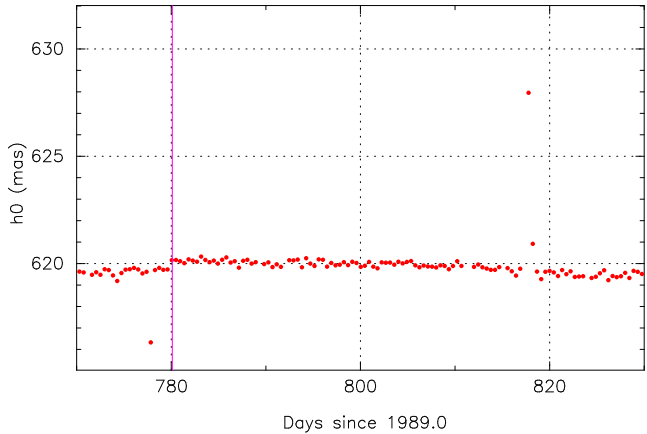}
\caption[Basic-angle calibration]{Basic-angle calibration values over a period
of two months. For the general trend the basic-angle variations are 
well below 1~mas, but at two instances deviating values are observed. Both
are directly related to known payload anomalies (new reduction). The vertical 
line at day 780 represents a refocusing of the instrument.
}
\label{fig:ba_calib}
\end{figure}
\begin{figure}
\centerline{\includegraphics[width=8.5cm]{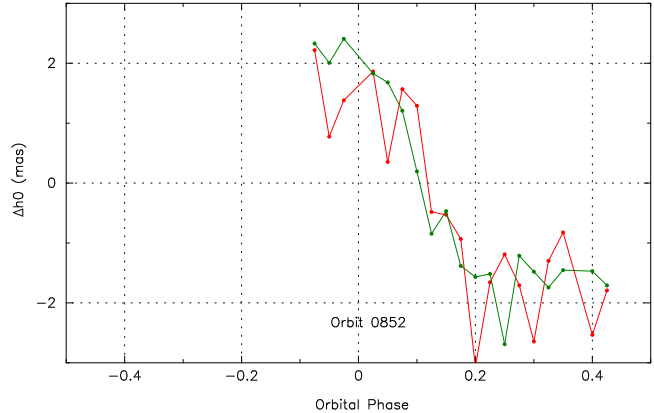}}
\caption[BA drift for orbit 852]{Drifting of the basic angle as observed in 
orbit 852, probably due to a restart of the on-board computer. The 
mean residuals over intervals of about 15 minutes are shown for the two 
fields of view, but for one of the two the sign was changed. Thus, what is 
shown is twice the correction to half the basic angle. In this orbit the 
basic angle correction drifted by 8 mas over a period of 5 hours.}
\label{fig:BasAng0852}
\end{figure}

Calibration of the basic angle was performed whenever sufficient data was
available. In general the basic angle shows only a slow evolution with time,
but on a number of occasions significantly deviating values are observed.
Both phenomena can be seen in Fig.~\ref{fig:ba_calib}, where outlying
points are observed for orbit 1060 (day 778) and orbits 1150 and 1151, day 
818. The first of these is an example of the most common reason for 
discrepant results: a restart of the on-board computer. These restarts
were necessitated by telemetry problems, and caused the entire payload,
including heater control, to be switched off. By the time control was restored,
the payload had changed temperature, which reflected in a change of the 
basic angle. Usually, within a few hours the temperature and basic angle value 
were back at their nominal values. Another example of this kind of event is 
shown in 
Fig.~\ref{fig:BasAng0852} for orbit 852 (day 686). A general characteristic 
of orbits thus affected is that a large fraction of the orbit, usually 
two rotations of the satellite or four hours, was spent on the recovery 
procedure, and did not produce observations.
The second event visible in Fig.~\ref{fig:ba_calib} is due to a one-off
anomaly, which was referred to in the operations report as an 
``anomalous under voltage''.

Drifts of the basic angle correction $h_0$ of up to 20~mas have been observed 
over the mission, and a total of 18 orbits are affected, 6 of these only 
marginally. Most of these events remained unnoticed in the production of the 
Hipparcos catalogue: the first-look processing \citep{schri85} focused 
primarily on well-covered orbits.

\begin{figure}
\centerline{\includegraphics[width=8.5cm]{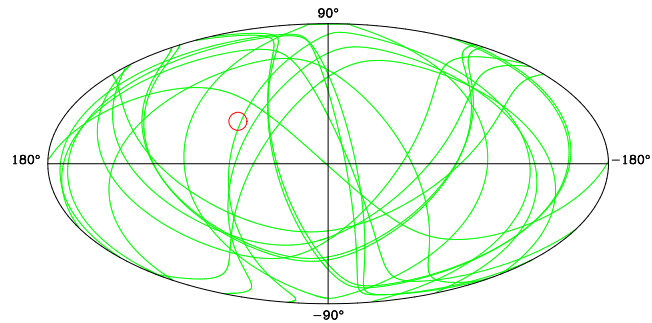}}
\caption{The distribution of orbits, in equatorial coordinates, that are 
affected by basic angle drift. The small circle shows the position and size
of the Pleiades field.}
\label{fig:DriftOrb}
\end{figure}
The new reduction of the data allows further systematic checks to be made
on the stability of the basic angle. These checks show generally a noise level
that is below 0.2~mas. At that level it becomes impossible to distinguish 
real basic-angle variations from remaining systematics in the astrometric 
catalogue. In the new reduction, basic-angle drift corrections are only 
applied for orbits with clearly recognisable drift problems.

\subsection{Conclusions on the basic-angle stability}

The basic-angle stability for the Hipparcos mission was better than the 
requirements, except for fewer than one per~cent of the orbits, most of which 
cover only half an orbit in data. The amount of data concerned is therefore 
very small, though for an individual star this does not need to be the case.
The map of the great circles with basic angle drift shows some areas 
where up to 5 of these circles cross close together (Fig.~\ref{fig:DriftOrb}). 
Correction of basic-angle drifts is often possible, 
though some loss of data quality is unavoidable for the most seriously
affected orbits, as in these cases also the scale in the instrument parameters
and the modulation parameters for the main grid are significantly 
affected.

\section{Connectivity}
\label{sec:conn} 
The second requirement for the construction of absolute parallaxes,
good connectivity between observations in the two fields of view, was 
generally assumed to have been taken care off by the construction of the 
Input Catalogue \citep{esa92} and the observing strategy. As was stated in 
the introduction, connectivity is obtained through the reconstruction of 
the along-scan attitude or scan phase. If the along-scan attitude 
reconstruction is based 
on one field of view only, it can adapt itself to ``faulty'' astrometry, 
such as a local offset of the parallax zero point. Only by ensuring that 
this reconstruction is based on (significant) contributions from observations 
in both fields of view it becomes possible to iterate out these local 
adjustments and obtain absolute parallaxes. The global tests on the parallax 
zero point and formal errors in the catalogue \citep{llind95, areno95} 
indicate that this condition was largely fulfilled for most of the sky in the 
preparation of the Hipparcos catalogue, but there may have been exceptions 
in specific regions with a high density of bright stars, such as some open 
cluster centres \citep{fvl05}.

\subsection{The Input Catalogue}
\label{sec:HIC}
The distribution of stars on the sky is not ideal for a mission like
Hipparcos, given the high range of stellar densities even for its limiting
magnitude (Hp=12). Limitations on the telemetry and
as imposed by the observing technique meant that no more than around 120~000
stars could be observed by the mission. These 120~000 stars had to be selected
from 214~000 stars requested in observing proposals. This task was 
assigned to the selection committee (led by Adriaan Blaauw) and the Input 
Catalogue consortium (led by Catherine Turon) \citep{esa89,esa92}.
The distribution of the requested stars reflects the main features of
stellar density on the sky, such as the Milky Way and open clusters. 
The highest densities for the requested stars, as measured over 6.4 square 
degree fields, are about 50 times higher than the lowest densities. After 
selection the typical density range in the Input Catalogue is about a 
factor 10. The final selection consists of two parts \citep{turon92}:
\begin{itemize}
\item A survey of about 52~000 bright stars, which is complete to a limiting 
magnitude as a function of spectral type and galactic latitude;
\item About 66~000 faint stars selected from the proposed observing programs.
\end{itemize}
The final selections were subjected to simulations to ensure that the expected
astrometric accuracies were possible to obtain 
\citep[Cr{\'e}z{\'e} in ][]{esa89}.

The selection was optimised on stellar density but not on integrated
light. The remaining intensity contrasts on the sky are generally
larger, as densely populated areas are often also areas
with bright stars. The rigidity of the Hipparcos catalogue depends on how 
the reduction software has handled those contrasts, which in turn is
determined by the formal errors assigned to observations in the great-circle
reduction process. The input data for this process are the reduced frame
transits, the grid-transit data collected for a star over a fixed 2.133~s 
period (roughly one ninth of a complete field-of-view transit). The 
modulation-phase estimates obtained for those transits have formal errors 
that reflect the signal modulation amplitude and Poisson statistics of the 
photon counts \citep[see also Fig.9 in ][]{paper4}. The only exceptions are 
the very brightest
stars (Hp$<2.5$) where saturation disturbed the statistics. The relation 
between the formal error on the phase estimate and the integrated intensity 
for a transit is nearly constant for the mission. The main dependence is the 
relative amplitude of the first harmonic in the grid-modulated signal. The 
formal errors on the transit positions are given by:
\begin{equation}
\sigma_a\approx \frac{255}{M1\sqrt{I_\mathrm{tot}}}~~~\mathrm{mas},
\label{eq:photnoise}
\end{equation}
where $I_\mathrm{tot}$ is the total photon count recorded for the transit, and
$M1$ the relative amplitude of the first harmonic in the signal modulation.
The main dependences of $M1$ are the focal adjustment of the telescope and the
colour index of the observed star. The value of $M1$ generally varied between
0.6 (red stars) and 0.8 (blue stars). 
 
For single frame transits $I_\mathrm{tot}$ ranges from a low of around 40 
up to $10^6$ for the brightest stars. 
\begin{figure}
\centerline{\includegraphics[width=8.5truecm]{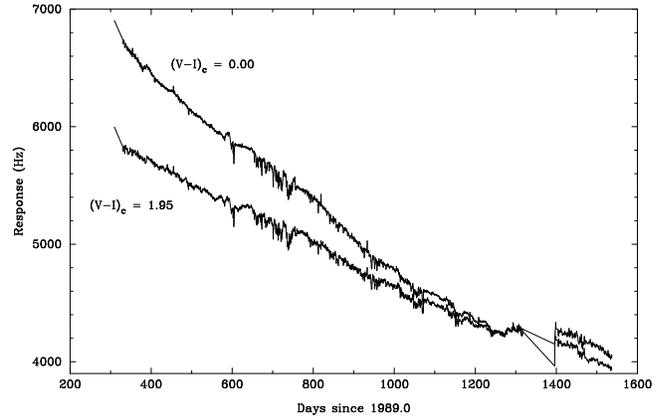}}
\caption{Time variation of the response in the centre of the field for 
8th mag stars of two different colour indices, showing the effect of aging of 
the detector chain through the decrease in response over the mission and 
the difference in this decrease for stars of different colour.}
\label{fig:itransm}
\end{figure}
Equation~\ref{eq:photnoise} provides the photon-statistics limit on the 
formal errors which ideally propagates all the way to the final astrometry. 
When also using the second harmonic, as was done by the FAST consortium, the 
scaling coefficient is
about 5~per~cent smaller: the second-order amplitude is approximately 
0.3 times the first harmonic amplitude, adding in weight 10~per~cent
to the solution, and reducing resulting formal errors by 5~per~cent. These
figures have been confirmed through extensive tests on the real data in the 
new reduction.  Thus, when using the first and second harmonic, as was done 
in the FAST reductions, the factor in Eq.~\ref{eq:photnoise} becomes 
$\approx242$~mas.

When the frame transits were used as input to the great-circle reduction, an 
assumed and fixed attitude noise was added in quadrature. The attitude
noise at the frame transit level is generally around 3~mas. This means 
that above a total count of $10^4$ the attitude noise is dominant. In 
comparison, the response of an 8$^\mathrm{th}$ magnitude star over the 
mission is shown in Fig.~\ref{fig:itransm}. Thus, from approximately 
magnitude 6--7 the measurement noise as implemented for the great-circle 
reduction was the fixed attitude noise rather than the photon-count 
statistics. This reduced the effective total intensity contrasts between 
two fields of view in some cases considerably. 

\begin{figure}
	\includegraphics[width=8.5cm]{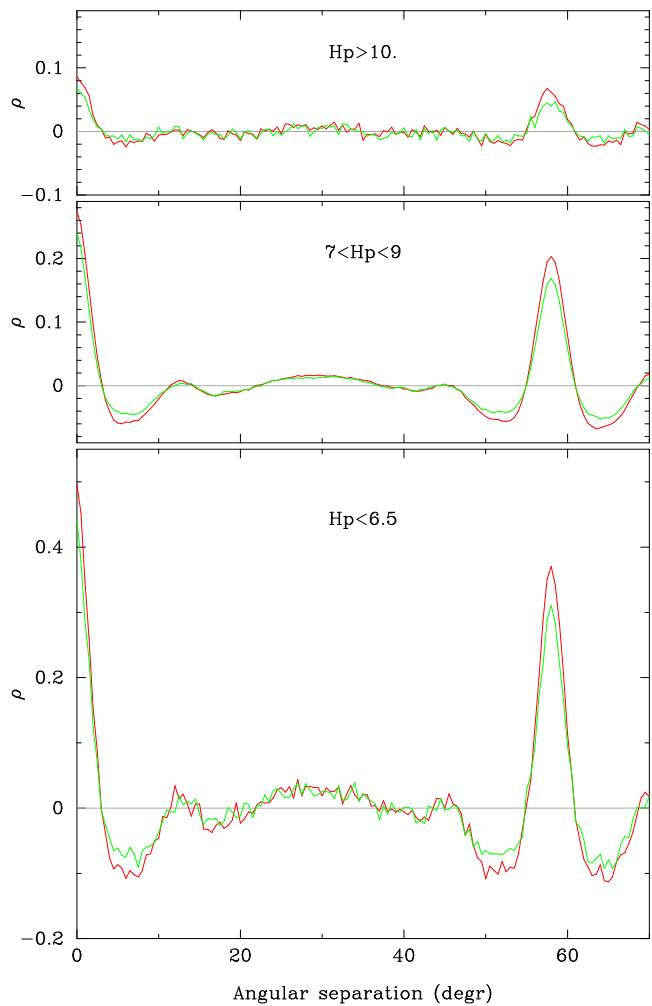}
	\caption{The zero- and first-order abscissa-error correlation
	peaks in the NDAC and FAST reductions for average length data sets
	and different magnitude ranges. The first-order peak shows the 
	correlations introduced through the simultaneous  measuring of stars 
	from the two fields of view. The zero-order peak shows these
	correlations for stars at small separations. These correlations
	are the result of attitude errors, which reflect most strongly for
	bright stars for which the photon-noise contribution is relatively 
	small.}
	\label{fig_corr}
\end{figure}

The connectivity of the data and the contribution of the attitude noise show
clearly from the abscissa-residual correlation statistics at orbit level. 
Average values for the whole mission have been presented in Volume~3 of
\citet{esa97} and by \citet{vLDWE}. However, averaged values 
are of limited use: the correlated errors originate from systematic errors 
in the along-scan attitude fitting, and these errors are much more dominant 
for bright stars than for faint ones (Fig.~\ref{fig_corr}).
Resolving the correlation statistics according to the magnitudes of the stars
involved is not possible in detail, but an impression of correlation 
statistics when both stars fall in a given magnitude interval can be obtained.
As one should expect, the correlations are considerably stronger for the
bright-star pairs than for faint-star pairs. Tests with mixed magnitude
pairs showed that the correlation statistics for these pairs is well 
represented by the statistics for the mean magnitude. The fact that 
significant correlations can still be observed even for stars of magnitude 10 
and above is probably a reflection of the unresolved scan-phase jumps and hits 
in the published data, as it implies the presence of some systematic errors at 
a level of 5~mas and more. Applying the adjusted correlation statistics to the 
Pleiades parallax solution provides a new cluster-parallax estimate of 
approximately 8.0~mas \citep{fvl05}, about halfway between the earlier 
Hipparcos-based estimate and the ``expected'' value, and with a difference 
at about 1~sigma of the formal error on the newly derived Hipparcos-based 
parallax.  

\subsection{The weight contrast in the Input Catalogue}
\label{sec:hotspots}

\begin{figure}
	\includegraphics[width=7.5cm]{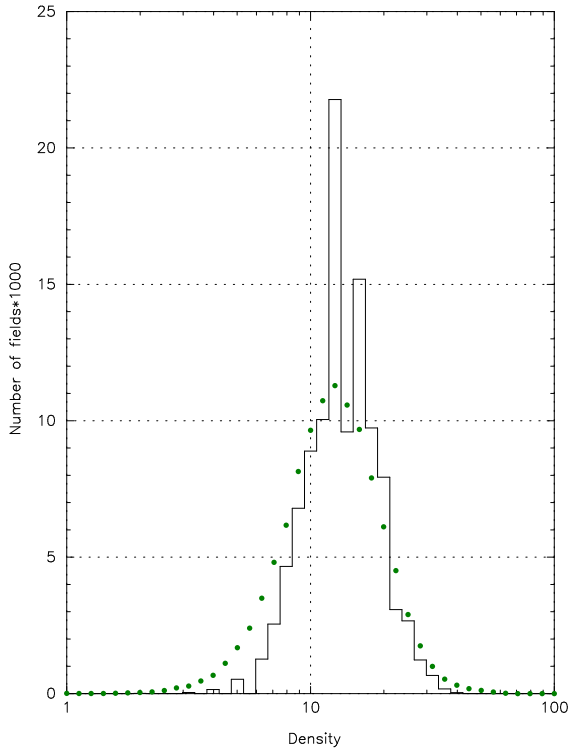}
	\caption{Histograms of the number of stars (line) and total intensity
	(dots) over 5.3 square-degrees fields centred on catalogue entries.
        The dynamic range (given by the width of the histogram) of the total 
	intensity per field is significantly higher than for the total number 
	of stars.}
	\label{fig:dens}
\end{figure}

The typical weight of a given area on the sky, in comparison with other 
areas, gives a more accurate impression of the remaining contrasts in the
Input Catalogue than comparing number densities of stars. These weights
determine the relative influence an area of the sky can exert on the along-scan
attitude determination. Assuming photon noise
and attitude noise as described above, the intensities of stars are added
rather than their numbers. The procedure is as follows. The Hp magnitudes
of all catalogue stars within a 1\fdg3 radius around each catalogue
star are collected and transformed to pseudo intensities. To represent the
modelling noise (mainly from the along-scan attitude reconstruction) all 
stars brighter than magnitude Hp$=7$ are treated as magnitude 7 stars. This
is further referred to as the minimum magnitude. The intensities are added 
to obtain an
integrated intensity. Thus, for each star we have available the number of
stars within a 1\fdg3 radius, and the integrated intensity for that area.
While the number counts have a dynamic range of about 20, the intensities
have a range of about 60 (Fig.~\ref{fig:dens}). The highest intensities are 
found for the cores of open clusters, in particular NGC~3114 in Carina and 
the Pleiades. Lowering the attitude noise, as is the case in the new 
reductions, lowers the minimum magnitude and further increases the contrast. 

Within the principles of the Hipparcos mission more weight is not necessarily
a good thing: areas with relatively high weight may not conform to the 
requirement of connectivity. The result can be that in such areas the 
fitting of the along-scan attitude is based almost entirely on the data from
that one field of view, preventing it from proper attachment to the rest of the
catalogue. This can lead locally to astrometric data for which the accuracy
is not fully reliable. An apparently discrepant parallax determination, as 
has been suggested to be the case for the Pleiades, could be the result. The 
astrometry in these fields may be partially
detached from the catalogue. As these fields rarely also contain stars at
sufficiently large distances to be used as check on the parallax zero point, 
it is difficult to verify independently whether this is or is not a 
serious (local) problem in the Hipparcos data. There is one indication, 
however, that it possibly is not. It would be expected, given the independent
reduction chains, for the results of such detached areas in the FAST and 
NDAC reductions to differ by more than the expected amount. When comparing
a map of the parallax differences between the FAST and NDAC results with
a map of the areas with high weight, there are at most a few marginal 
correlations. These comparisons indicate that the detachment due to excess 
weight may not play a very serious role in the reliability of the Hipparcos 
catalogue. There is one proviso that needs to be made here: the two consortia
started reductions using the same Input Catalogue \citep{esa92}, and used 
nearly the same data (the only differences being in stretches of data accepted 
or rejected). Local systematic errors due to detachment may possibly, as a 
result, have developed in a similar way.  

\subsection{Is re-attachment of detached areas possible?}
\label{sec:correct}
It is to the credit of Valeri Makarov who first identified some of 
the effects described above as a possible reason for the problems experienced
with the Pleiades parallax. Makarov also suggested and applied a method for
correcting (re-connecting?) detached areas \citep{makar02,makar03}. A range 
of assumptions is made by Makarov about the way observations are distributed
between the two fields of view. The availability of the raw data and a 
complete data analysis package allows checking those assumptions. The method
of Makarov is based on the idea of incorporating the abscissa residuals
from the other field of view to correct those of a potentially detached
area. This idea had been first put forward by \citet{fvl99b}, but its
implementation was rejected on statistical grounds: for the field covering
the centre of the Pleiades cluster the unit-weight variance 
of the potentially coinciding abscissa residuals was observed to be smaller
than that of the non-coinciding residuals. 

Makarov accumulated the 
data from great circles on which observations of a target area are contained, 
and selected the data from stars positioned on those circles at $58\pm0.7$ 
degrees from the target field, the so-called coinciding stars. By assuming 
that all these coinciding stars had half of their observations coinciding 
with the target field, and half not, he 
argued that any systematic residual caused by the target field will on average
equal half of the residual of the coinciding star for that circle. This
translates into a correction coefficient $R=2$. 

To check the validity of this correction coefficient, and using the raw data, 
a total of 1400 great circles were investigated, each containing the transits 
of at least 1000 different stars. For each star the potentially coinciding 
stars were selected, where a potentially coinciding star is situated at
$58\pm0.7$ degrees from the target star. Both the target star and the 
coinciding star have potentially observations in both fields of view, and
in only one of those FOVs the measurements may coincide. This situation is 
represented by a statistical parameter $p$. If there are in total $n_t$
observations for the target star, and $n_c$ observations for the coinciding
star, and there are $n_k$ coinciding observations, then:
\begin{equation}
p \equiv n_k / \sqrt{n_t\cdot n_c}
\end{equation}  
In an ideal situation, both stars have been observed the same number of 
times, equally distributed between the two FOVs, and half of these
observations coincided with the other star. In that case, $p=0.5$. This is
equivalent to setting $R=2$ in \citet{makar02}. The data for several 
million actual field transits shows, however, that the real situation is 
far from ideal. 
\begin{figure}
	\includegraphics[width=8.5cm]{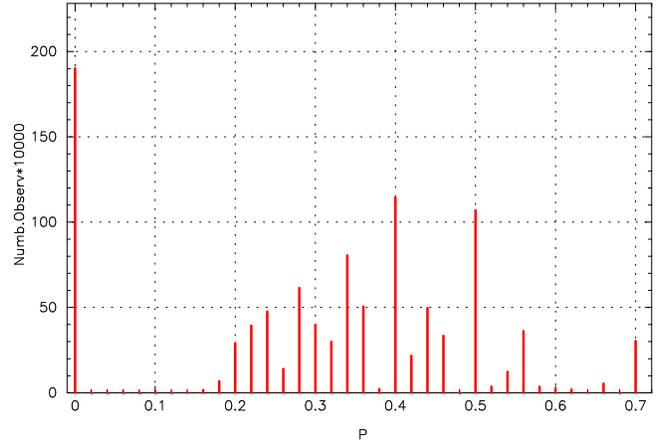}
	\caption{Distribution of the coincidence statistic $p$ for a
	large selection of stars, restricted to an ordinate difference of 
	less than 0.75~degrees. The peak at $p=0$ represents the unsuccessful
	potentially-coinciding stars.}
	\label{fig_coin}
\end{figure}
To start, 35 per~cent of the potentially coinciding stars
never have any coinciding measurements and can therefore be only 
indirectly correlated. Most of these missed chances are due to the 
ordinate differences between the target and the coinciding star. By 
limiting the ordinate difference to 0.75 degrees, only 13 per~cent does
not coincide. But for those that do coincide, the distribution over the 
statistic $p$ shows an efficiency that is well below the ideal value of
0.5 (see Fig.~\ref{fig_coin}). The average value for $p$ in this selection 
is 0.32 (without the additional selection it is 0.26). The average
value of $p$ will decrease with increasing length of the time interval
covered by the great circle, as is also reflected in the correlation
statistics by \citet{vLDWE}. Taking therefore all potentially coinciding
stars as actually coinciding, and at the ideal distribution of observing
time, seems highly unjustified.

The use of the abscissa residuals from coinciding transits has a further
risk. The noise on those residuals is composed of two main contributions:
Poisson and attitude noise. As was explained above, Poisson noise dominates 
for stars fainter than Hp$=8$, while attitude noise dominates for stars 
brighter than Hp$=6$. Transferring abscissa residuals from faint 
coinciding stars onto the target field will not be of any benefit for the
target field, it will only increase the overall noise levels. The fact
that a possibly more likely answer is found by Makarov using this method
is no proof of the validity of the method. In fact, his method contradicts his 
probably correct assumption about the problem with the Pleiades parallax: 
when assuming that the Pleiades region is detached from the catalogue, it is
not possible to correct it with methods that would apply only under condition
of (nearly) ideal connectivity. On the other hand, if areas are ideally 
connected then there is no reason to apply a correction.

\subsection{Conclusions on connectivity}
Connectivity may have been a relatively minor issue in the construction of
the Hipparcos catalogue, affecting only a few extreme areas of the sky.
Unfortunately, these areas also tend to be some of the most interesting.
However, when in a new reduction the attitude noise contribution is 
considerably reduced, the weight contrasts will increase and many more areas
will be affected. This can be prevented by artificially damping the
weight ratio between the two fields of view to a reasonable fraction. Even 
so it will still be necessary to iterate the solution to ensure full
connectivity over the whole sky. Any effects of poor connectivity in the
published catalogue cannot be corrected afterwards from the intermediate 
astrometric data. Connectivity is very relevant for the reduction of the
astrometric data of the Gaia mission \citep{perry02}, where contrasts 
are purely set by the actual stellar density variations on the sky, and reach
much higher values than encountered for the Hipparcos mission. 
 
\section{Some statistical properties resulting from the merging of the data}
\label{sec:merg}
The great-circle reduction process and the associated potential for
instabilities made a clean propagation of formal errors virtually 
impossible. In addition, the total integrated intensities of transits
were not propagated through the reduction chain, thus a handle on those formal
errors was effectively lost. This led to various ``adjustments'' of 
errors along the way: in the great-circle reduction results, the sphere
reconstruction and the data merging. Some adjustments were made as a 
function of stellar magnitude, while observing-time variations implied
that there was no clean direct relation between formal errors and those
magnitudes. Most, but not all, features observed in the formal errors
on the published (per-orbit) abscissae can be understood from the descriptions
in Volume 3, Chapters 9, 11, 16 and 17 of \citet{esa97}.   
\subsection{Propagation of errors}
\label{sec:prop_errors}
As was stated before, the errors on the Hipparcos astrometry all originate 
from two sources: 
photon noise and modelling errors. The photon noise is a function of the 
integrated photon-count and the modulation amplitude $M1$, and reflects 
in the formal errors on the positional estimates as described by 
Eq.~\ref{eq:photnoise}. 

\begin{figure}
	\includegraphics[width=8.5cm]{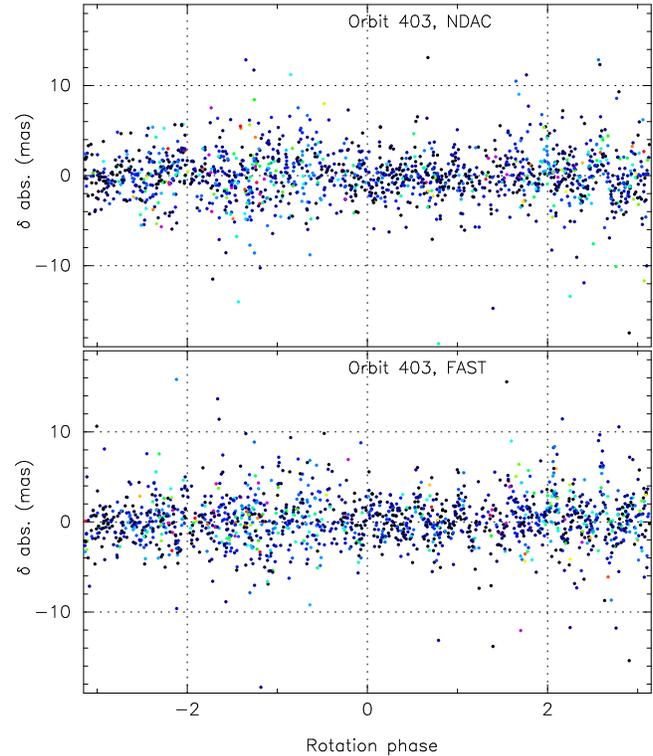}
	\caption{Abscissa residuals for the FAST and NDAC reductions of 
	orbit 403. Some remaining systematics are still visible in both 
	graphs.}
	\label{fig:resid0403}
\end{figure}
\begin{figure}
	\includegraphics[width=8.5cm]{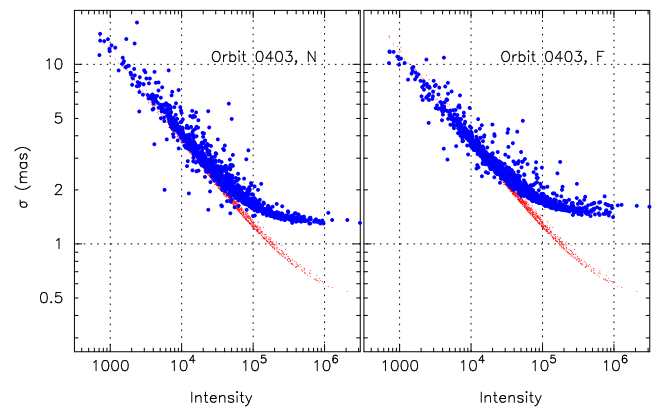}
	\caption{Formal errors as assigned to the abscissa residuals for
	orbit 403. The thin dots in the background give the formal
	errors for the new reduction, as based on the transit intensities
	and modulation amplitude of the first harmonic, $M1$.}
	\label{fig:ferr0403}
\end{figure}
The modelling noise consists of two main contributions: reconstruction of the 
along-scan phase (attitude) and reconstruction of the instrument parameters,
describing the differential relations between positions on the sky and on 
the grid for the two fields of view and for stars of different colour indices.
The first of these two generally dominates the modelling noise, and covers 
all the problems presented earlier in this paper: phase jumps, hits, 
eclipses and solution instabilities. 

\subsection{Statistical peculiarities of the merged data}
\label{sec:StatPec}
The level of the attitude noise was set empirically, based on the observed 
variance of the abscissa residuals following the great-circle 
reduction. The attitude noise thus introduced varies strongly from 
data set to data set, between lows of 1.3~mas and highs of 10~mas. The
high noise values always hide strong systematics in the residuals.
An example of a typical good data set is shown in Fig.~\ref{fig:resid0403}
and Fig.~\ref{fig:ferr0403}, with an attitude noise level for NDAC of
1.3~mas and for FAST 1.5~mas. The photon noise is observed close to the 
expected relation, slightly lower for FAST than for NDAC due to the inclusion
of the second harmonic.

\begin{figure}
	\includegraphics[width=8.5cm]{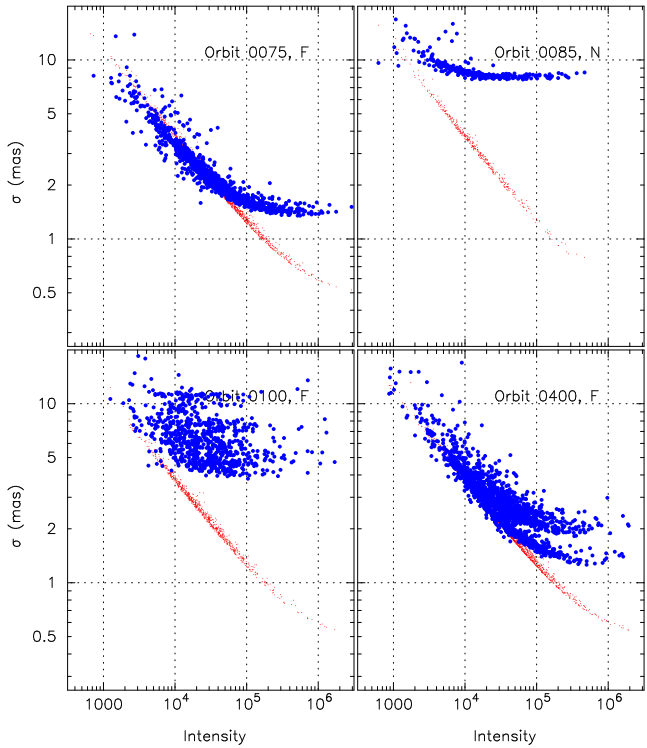}
	\caption{Peculiarities in the formal errors on abscissa residuals 
	  for some selected orbits (see text for details). N: NDAC, F: FAST}
	\label{fig:ferranom}
\end{figure}
Several deviations from this more or less ideal situation are observed.
Examples of some typical situations are shown in Fig.~\ref{fig:ferranom}.
The first example, orbit 75, top left, shows the effects of a single 
correction applied to the formal errors as part of the catalogue merging. 
As part of the merging process, the standard errors on the FAST and NDAC
great-circle results were ``normalised''. However, as at that stage 
information on the two noise components was no longer available, the
``normalisation'' was applied to all observations. As a result, it appears
that noise levels significantly below the minimum photon noise are assigned 
to the fainter transits, which seems unrealistic. This situation occurs
frequently for data sets from the first few months of the mission, as 
reduced by FAST. 

The second example, orbit 85, top right, shows an extreme case of the attitude 
noise contribution. In this case the attitude noise hides large-scale
systematics in the abscissa residuals for a short stretch of data, caused
by the instabilities in the great-circle reduction described above. The 
attitude noise in these data sets is highly non-Gaussian.

The third example, orbit 100, bottom left, shows a situation that looks like
multiple attitude noise levels having been applied to various short 
stretches of data. The cause of this behaviour is still unclear. It
occurs, to various degrees of severity, in quite a few FAST data sets.

The final example, orbit 400, bottom right, shows a case where in the FAST 
reduction an orbit was split in two parts because of the presence of 
a large data gap (in this case the gap was just over one revolution of the
satellite). Each part of the reduction was assigned an attitude noise level, 
and the two parts were later merged to one data set in the preparations for
the merging of the FAST and NDAC data.

\begin{figure}
\centerline{\includegraphics[width=8.5cm]{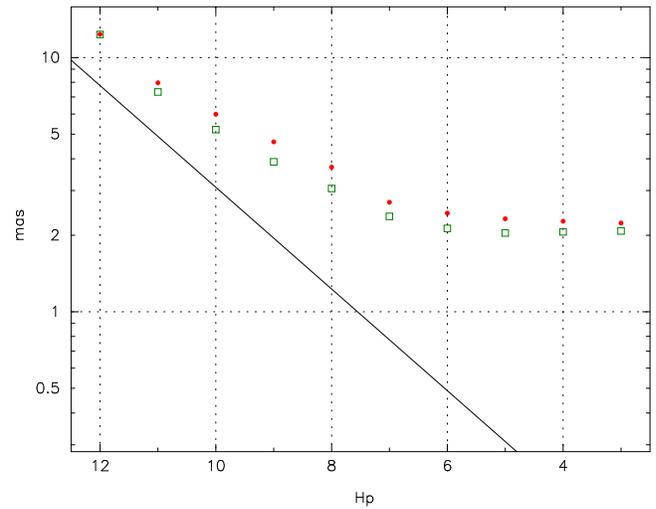}}
\caption{Standard deviations ($\sigma$ in mas) for the abscissa residuals
as a function of the $\mathrm{H}p$ magnitude. Open symbols refer to FAST 
data, filled symbols to NDAC data. The diagonal line represents the expected 
relation for Poisson noise on the photon counts only, given identical 
integration times for all magnitudes. The flattening of the relation towards 
brighter magnitudes shows the 2~mas noise contribution from the instrument 
parameters and attitude modelling.}
\label{fig:sd_absc_err}
\end{figure}
How these formal errors translated into dispersions of abscissa residuals
as a function of magnitude is shown in Fig.~\ref{fig:sd_absc_err}. The
initial departure from the photon-noise relation is mainly due to 
amounts of observing time assigned to stars of different magnitudes. The 
flattening towards the brighter stars shows the average attitude noise at
orbit-transit level for the published data.

These statistical properties of the abscissa data imply that it is close
to impossible to provide a reliable global abscissa-error correlation 
function for the Hipparcos data. Conditions differ from one orbit to the 
next. The same is probably also true of the correlation coefficients
between the FAST and NDAC data. As a result, studies of astrometric
parameters for open clusters are severely hampered by intrinsic uncertainties,
which can no longer be solved \textit{a posteriori} from the published data.
The only reasonable way forward appears to be a complete re-reduction of 
the raw Hipparcos data, without the use of the great-circle reduction, 
but starting from the current published catalogue. Only that way can 
large-scale systematics and the associated abscissa error correlations 
as well as anomalous formal errors on the abscissae be eliminated.  

\section{Statistical verification of the astrometric data}
\label{sec:stat_astr}
The verification of a data set claiming a unique level of accuracy
is always very difficult. For the Hipparcos data there were available
a very limited number of parallaxes with compatible accuracies, and
proper motions based on a very much longer base line than explored by 
Hipparcos. Positional reference frames in optical wavelengths as obtained with
ground-based instrumentation suffer from systematic errors well beyond the
accuracy level of the Hipparcos data. There is, however, a small number of
radio sources with astrometric parameters of compatible accuracy.

Several tests have been applied to the Hipparcos data to test its reliability:
\begin{itemize}
\item Comparisons with any available astrometric data of compatible
accuracy, in optical and radio wavelengths;
\item Testing of formal errors and parallaxes for objects with 
negligible parallaxes, such as stars in the SMC and LMC;
\item Comparing the distribution of negative parallaxes with the 
expected distribution for the given formal errors;   
\item Comparing globally parallaxes for open clusters with pre-Hipparcos
estimates of their distances.
\end{itemize}
These tests were carried out by Lennart Lindegren and Fr{\'e}d{\'e}ric
Arenou \citep{llind95,areno95}, and indicated quite clearly that globally
the parallaxes and formal errors formed a reliable and internally consistent
data set. The independent tests by Arenou and Lindegren showed no 
systematic bias in the parallax zero point down to a level of 0.1~mas, and
confirmed the formal errors to agree with actual errors to within 2 to 3
per~cent.

There are some aspects of these tests that would warrant caution on 
their results. The first concerns the magnitude range (and thus the formal 
error range) for the tests. The tests are dominated by data for faint stars,
with errors set by photon noise only. Bright stars are generally only 
suitable for comparisons with ground-based observations which tend to 
be of much lower accuracy than the Hipparcos data. As was shown above, 
potential connectivity problems are in particular relevant for the
brighter stars. 

There is also still a small problem with the proper motions of
cluster members. Internal proper motion dispersions in open clusters
have been studied on the basis of differential astrometry with 10 times
higher precision than the Hipparcos data \citep[see for example][]{fvl94}.
For the core of the Pleiades this dispersion is about 0.8~mas~\peryr. 
The internal dispersions in the open cluster proper motions as derived by
\citet{fvl99a} and \citet{robic99} from the Hipparcos data on the basis of 
the overall dispersion of the proper motions and their formal errors, are 
at a level of 1.5 to 2~mas~\peryr, well above the expected 
values. Detailed problems with the proper motions of cluster stars are also 
indicated when comparing the differential components of the Pleiades proper 
motions as measured by Hipparcos with the best available differential 
determination by \citet{vasil79}. Given the distances of the clusters 
involved, this is unlikely to be an effect of unresolved orbital motions.

\section{Conclusions}
\label{sec:concl}

Above all, it should be recognised that the two data reduction consortia,
in close collaboration with the ESA teams at ESOC and ESTEC, produced 
a Hipparcos catalogue with astrometric accuracies well exceeding the 
original aims and expectations of the mission, and did so under the very
difficult conditions set by the orbit anomaly. Criticisms that have been 
expressed, concern the reliability of the catalogue for one or two 
small areas of the sky, where conditions for reconstructing the best possible
astrometry may not have been sufficiently well understood during the 
data reductions. These aspects, and a range of other partly overlooked 
disturbances of the data, as well as the possibility of now sidestepping the
great-circle reduction process, with its potential instabilities, still 
create opportunities for significant further improvements to the catalogue, 
in both overall accuracy and statistical reliability. As the Hipparcos data 
is irreplaceable, it is felt that this new reduction of the Hipparcos data 
needs to be done to preserve the best possible results from what has so far 
been a unique experiment. Furthermore, the experience obtained with the 
new analysis and reduction of the Hipparcos data is of direct benefit for
the forthcoming Gaia mission, in particular concerning overall connectivity 
for the reconstructed astrometric catalogue and the importance of rigidity of 
the spacecraft for the along-scan attitude reconstruction.

\begin{acknowledgements}
It is a pleasure to thank Elena Fantino for her contributions to the current 
study. The comments of Dafydd W.~Evans, Robin Catchpole, Rudolf Le Poole and
Anthony Brown on earlier versions of the manuscript, as well as suggestions
from the referee, Ulrich Bastian, are also much appreciated.  
\end{acknowledgements}
%\begin{figure}
%\resizebox{0.75\textwidth}{!}{%
%  \includegraphics{example.eps}
%\vspace{5cm}       % Give the correct figure height in cm
%\end{figure}
%\begin{figure*}
%\vspace*{5cm}       % Give the correct figure height in cm
%\end{figure*}
%\begin{table}
%\begin{tabular}{lll}
%\hline\noalign{\smallskip}
%%\end{tabular}
% Or use
%\vspace*{5cm}  % with the correct table height
%\end{table}
%
% BibTeX users please use
\bibliographystyle{aa}
\bibliography{3192}

\begin{thebibliography}{49}
\expandafter\ifx\csname natexlab\endcsname\relax\def\natexlab#1{#1}\fi

\bibitem[{Arenou {et~al.}(1995)Arenou, Lindegren, Fr{\oe}schl\'e, G\'omez,
  Turon, Perryman, \& Wielen}]{areno95}
Arenou, F., Lindegren, L., Fr{\oe}schl\'e, M., {et~al.} 1995, \aap, 304, 52

\bibitem[{Castellani {et~al.}(2002)Castellani, Degl'Innocenti, Moroni, \&
  Tordiglione}]{castel02}
Castellani, V., Degl'Innocenti, S., Moroni, P. G.~P., \& Tordiglione, V. 2002,
  \mnras, 334, 193

\bibitem[{{Dalla Torre} \& van Leeuwen(2003)}]{paper1}
{Dalla Torre}, A. \& van Leeuwen, F. 2003, Space Sci.Rev., 108, 451

\bibitem[{Donati \& Sechi(1992)}]{donat92}
Donati, F. \& Sechi, G. 1992, \aap, 258, 46

\bibitem[{ESA(1989)}]{esa89}
ESA, ed. 1989, The Hipparcos Mission, SP No. 1111 (ESA)

\bibitem[{ESA(1992)}]{esa92}
ESA, ed. 1992, The Hipparcos Input Catalogue, SP No. 1136 (ESA)

\bibitem[{ESA(1997)}]{esa97}
ESA, ed. 1997, The Hipparcos and Tycho Catalogues, SP No. 1200 (ESA)

\bibitem[{{Knapp} {et~al.}(2001){Knapp}, {Pourbaix}, \& {Jorissen}}]{knapp01}
{Knapp}, G., {Pourbaix}, D., \& {Jorissen}, A. 2001, \aap, 371, 222

\bibitem[{{Knapp} {et~al.}(2003){Knapp}, {Pourbaix}, {Platais}, \&
  {Jorissen}}]{knapp03}
{Knapp}, G.~R., {Pourbaix}, D., {Platais}, I., \& {Jorissen}, A. 2003, \aap,
  403, 993

\bibitem[{Kovalevsky(1998)}]{koval98}
Kovalevsky, J. 1998, \araa, 36, 99

\bibitem[{{Kovalevsky} {et~al.}(1992){Kovalevsky}, {Falin}, {Pieplu},
  {Bernacca}, {Donati}, {Froeschle}, {Galligani}, {Mignard}, {Morando},
  {Perryman}, {Schrijver}, {van Daalen}, {van der Marel}, {Villenave},
  {Walter}, {Badiali}, {Borriello}, {Brouw}, {Canuto}, {Guerry}, {Hering},
  {Huc}, {Iorio-Fili}, {Lacroute}, {Lattanzi}, {Le Poole}, {Murgolo},
  {Preston}, {R{\" o}ser}, {Sanso}, {Wielen}, {Belforte}, {Bernstein},
  {Bucciarelli}, {Cardini}, {Emanuele}, {Fassino}, {Lenhardt}, {Lestrade},
  {Prezioso}, \& {Tommasini Montanari}}]{koval92}
{Kovalevsky}, J., {Falin}, J.~L., {Pieplu}, J.~L., {et~al.} 1992, \aap, 258, 7

\bibitem[{Lindegren(1995)}]{llind95}
Lindegren, L. 1995, \aap, 304, 61

\bibitem[{Lindegren {et~al.}(1992)Lindegren, H{\o}g, van Leeuwen, Murray,
  Evans, Penston, Perryman, Petersen, Ramamani, \& Snijders}]{lindeg92}
Lindegren, L., H{\o}g, E., van Leeuwen, F., {et~al.} 1992, \aap, 258, 18

\bibitem[{Makarov(2002)}]{makar02}
Makarov, V. 2002, \aj, 124, 3299

\bibitem[{{Makarov}(2003)}]{makar03}
{Makarov}, V.~V. 2003, \aj, 126, 2408

\bibitem[{{Munari} {et~al.}(2004){Munari}, {Dallaporta}, {Siviero}, {Soubiran},
  {Fiorucci}, \& {Girard}}]{munar04}
{Munari}, U., {Dallaporta}, S., {Siviero}, A., {et~al.} 2004, \aap, 418, L31

\bibitem[{Narayanan \& Gould(1999)}]{naray99}
Narayanan, V.~K. \& Gould, A. 1999, \apj, 523, 328

\bibitem[{Pan {et~al.}(2004)Pan, Shao, \& Kulkarni}]{pan04}
Pan, X., Shao, M., \& Kulkarni, S.~R. 2004, Nature, 427, 326

\bibitem[{{Percival} {et~al.}(2005){Percival}, {Salaris}, \&
  {Groenewegen}}]{perciv05}
{Percival}, S.~M., {Salaris}, M., \& {Groenewegen}, M.~A.~T. 2005, \aap, 429,
  887

\bibitem[{Percival {et~al.}(2003)Percival, Salaris, \& Kilkenny}]{perciv03}
Percival, S.~M., Salaris, M., \& Kilkenny, D. 2003, \aap, 400, 541

\bibitem[{{Perryman}(2002)}]{perry02}
{Perryman}, M.~A.~C. 2002, \apss, 280, 1

\bibitem[{{Perryman} {et~al.}(1997){Perryman}, {Lindegren}, {Kovalevsky},
  {H{{\o}}}, {Bastian}, {Bernacca}, {Cr{\' e}z{\' e}}, {Donati}, {Grenon}, {van
  Leeuwen}, {van der Marel}, {Mignard}, {Murray}, {Le Poole}, {Schrijver},
  {Turon}, {Arenou}, {Froeschl{\' e}}, \& {Petersen}}]{perry97L}
{Perryman}, M.~A.~C., {Lindegren}, L., {Kovalevsky}, J., {et~al.} 1997, \aap,
  323, L49

\bibitem[{Pinsonneault {et~al.}(1998)Pinsonneault, Staufer, Soderblom, King, \&
  Hanson}]{pinso98}
Pinsonneault, M.~H., Staufer, J., Soderblom, D.~R., King, J.~R., \& Hanson,
  R.~B. 1998, \apj, 504, 170

\bibitem[{Pinsonneault {et~al.}(2003)Pinsonneault, Terndrup, Hanson, \&
  Stauffer}]{pinso03}
Pinsonneault, M.~H., Terndrup, D.~M., Hanson, R.~B., \& Stauffer, J.~R. 2003,
  \apj, 598, 588

\bibitem[{Pinsonneault {et~al.}(2000)Pinsonneault, Terndrup, \& Yuan}]{pinso00}
Pinsonneault, M.~H., Terndrup, D.~M., \& Yuan, Y. 2000, in Stellar clusters and
  associations: convection, rotation and dynamos, ed. R.~Pallavicini,
  G.~Micela, \& S.~Sciortino, Vol. 198 (PASPC), 95

\bibitem[{{Pourbaix} \& {Boffin}(2003)}]{pourb03}
{Pourbaix}, D. \& {Boffin}, H.~M.~J. 2003, \aap, 398, 1163

\bibitem[{{Pourbaix} \& {Jorissen}(2000)}]{pourb00}
{Pourbaix}, D. \& {Jorissen}, A. 2000, \aaps, 145, 161

\bibitem[{{Raboud} \& {Mermilliod}(1998)}]{raboud98}
{Raboud}, D. \& {Mermilliod}, J.-C. 1998, \aap, 329, 101

\bibitem[{Reid(1999)}]{reid99}
Reid, I.~N. 1999, Ann.Rev.Astron.Astrophys., 37, 191

\bibitem[{Robichon {et~al.}(1999)Robichon, Arenou, Mermilliod, \&
  Turon}]{robic99}
Robichon, N., Arenou, F., Mermilliod, J.~C., \& Turon, C. 1999, \aap, 345, 471

\bibitem[{Schrijver(1985)}]{schri85}
Schrijver, J. 1985, in The second FAST thinkshop, ed. J.~Kovalevsky, 375--378

\bibitem[{{Soderblom} {et~al.}(2004){Soderblom}, {Benedict}, {Nelan},
  {McArthur}, {Spiesman}, {Ramirez}, {Jones}, {Pinsonneault}, {Torres},
  {Latham}, {Franz}, \& {Wasserman}}]{soder04}
{Soderblom}, D.~R., {Benedict}, G.~F., {Nelan}, E., {et~al.} 2004, American
  Astronomical Society Meeting, 204,

\bibitem[{Soderblom {et~al.}(1998)Soderblom, King, Hanson, Jones, Fischer, \&
  Stauffer}]{soder98}
Soderblom, D.~R., King, J.~R., Hanson, R.~B., {et~al.} 1998, \apj, 504, 192

\bibitem[{Stello \& Nissen(2001)}]{stello01}
Stello, D. \& Nissen, P.~E. 2001, \aap, 374, 105

\bibitem[{{Turon} {et~al.}(1992){Turon}, {Gomez}, {Crifo}, {Creze}, {Perryman},
  {Morin}, {Arenou}, {Nicolet}, {Chareton}, \& {Egret}}]{turon92}
{Turon}, C., {Gomez}, A., {Crifo}, F., {et~al.} 1992, \aap, 258, 74

\bibitem[{van~der Marel(1988)}]{vdmar88}
van~der Marel, H. 1988, PhD thesis, Technische Universiteit Delft

\bibitem[{van~der Marel \& Petersen(1992)}]{vdmar92}
van~der Marel, H. \& Petersen, C.~S. 1992, \aap, 258, 60

\bibitem[{{van Leeuwen}(1980)}]{fvl80}
{van Leeuwen}, F. 1980, in IAU Symp. 85: Star Formation, 157--162

\bibitem[{van Leeuwen(1994)}]{fvl94}
van Leeuwen, F. 1994, in Galactic and Solar System Optical Astrometry, ed.
  L.~V. Morisson \& G.~Gilmore (Cambridge University Press), 223

\bibitem[{van Leeuwen(1997)}]{fvl97}
van Leeuwen, F. 1997, Space Sci.Rev., 81, 201

\bibitem[{van Leeuwen(1999{\natexlab{a}})}]{fvl99a}
van Leeuwen, F. 1999{\natexlab{a}}, \aap, 341, L71

\bibitem[{van Leeuwen(1999{\natexlab{b}})}]{fvl99b}
van Leeuwen, F. 1999{\natexlab{b}}, in Harmonizing cosmic distance scales in a
  post-Hipparcos era, ed. D.~Egret \& A.~Heck, Vol. 167 (PASPC), 52--71

\bibitem[{van Leeuwen(2005)}]{fvl05}
van Leeuwen, F. 2005, in Transit of Venus: New views of the Solar System and
  Galaxy, ed. D.~Kurtz \& G.~Bromage (Cambridge University Press)

\bibitem[{van Leeuwen \& Evans(1998)}]{vLDWE}
van Leeuwen, F. \& Evans, D.~W. 1998, \aap, 323, 157

\bibitem[{{van Leeuwen} {et~al.}(1992){van Leeuwen}, {Evans}, {Lindegren},
  {Penston}, \& {Ramamani}}]{fvl92a}
{van Leeuwen}, F., {Evans}, D.~W., {Lindegren}, L., {Penston}, M.~J., \&
  {Ramamani}, N. 1992, \aap, 258, 119

\bibitem[{van Leeuwen \& Fantino(2003)}]{paper4}
van Leeuwen, F. \& Fantino, E. 2003, Space Sci.Rev., 108, 537

\bibitem[{van Leeuwen {et~al.}(1992)van Leeuwen, Penston, Perryman, Evans, \&
  Ramamani}]{fvl92b}
van Leeuwen, F., Penston, M.~J., Perryman, M. A.~C., Evans, D.~W., \& Ramamani,
  N. 1992, \aap, 258, 53

\bibitem[{{Vasilevskis} {et~al.}(1979){Vasilevskis}, {van Leeuwen},
  {Nicholson}, \& {Murray}}]{vasil79}
{Vasilevskis}, S., {van Leeuwen}, F., {Nicholson}, W., \& {Murray}, C.~A. 1979,
  \aaps, 37, 333

\bibitem[{{Zwahlen} {et~al.}(2004){Zwahlen}, {North}, {Debernardi}, {Eyer},
  {Galland}, {Groenewegen}, \& {Hummel}}]{zwahl04}
{Zwahlen}, N., {North}, P., {Debernardi}, Y., {et~al.} 2004, \aap, 425, L45

\end{thebibliography}
% \begin{thebibliography}{}
%Author, Journal \textbf{Volume,} (year) page numbers.
%Author, \textit{Book title} (Publisher, place year) page numbers
% etc
%\end{thebibliography}

\end{document}